\documentclass[a4paper, 11pt]{article}
\usepackage{indentfirst}
\usepackage{graphicx}
\usepackage{epstopdf}
\usepackage{epsfig}
\usepackage{overpic}
\usepackage{overcite}
\usepackage{array}
\usepackage{color}
\usepackage{multirow}
\usepackage{float}
\usepackage{subfigure}
\usepackage{amsmath,amssymb, amsthm}
\usepackage[mathscr]{euscript}
\usepackage{latexsym,bm}
\usepackage{cite}
\usepackage{booktabs}

\newtheorem{theorem}{Theorem}
\newtheorem{lemma}{Lemma}

\newtheorem{algorithm}{Algorithm}

\newcommand\comment[1]{}

\graphicspath{{images/}}
{\theoremstyle{remark} \newtheorem{remark}{Remark}}

\begin{document}
\begin{center}

{\LARGE\bf A trigonometric integrator for the constrained ring polymer Hamiltonian dynamics} {\vskip 0.5cm}Yunfeng Xiong \footnote{
Email addresses: xiongyf@zju.edu.cn.\\
.} \\
  \small (Department of Mathematics, Zhejiang University, Hangzhou 310027, Zhejiang, P.R.China)\\

 \end{center}{\vskip 0.5cm}


\begin{abstract} 
A class of trigonometric integrator is proposed for the constrained ring polymer Hamiltonian dynamics, arising from the path integral molecular dynamics. The integrator is formulated by the composition of flows, thereby integrating the Cartesian equations of motions under normal mode representation and preserving the holonomic constraints by iterations. It is illustrated that the trigonometric method can preserve the symplectic structure and time-reversibility, and its near-conservation of Hamiltonian is analyzed in the framework of modulated Fourier expansion analysis. Numerical examples illustrating its stability are presented using the SPC/E force field at 298K.
    
\vspace*{4mm}
\end{abstract}

\section{Introduction}
Molecular dynamics simulations and Ab initio calculations are powerful and important tools in modern computational chemistry\cite{LEC}. Classical MD simulations, neglecting the quantum effects, can deal with a wide range of experimental dynamics properties. Ab initio calculations, on the other hand, provide a more precise description of quantized particles, but solving the quantum dynamics of many-body systems remains one of the most challenging problems to the unfavorable computer scaling with system size and time scale.
 
The calculations of static equilibrium properties of a quantum mechanical system are comparatively easy by investigating the path integral representation. These methods, including the primitive path integral molecular dynamics (PIMD)\cite{DMPM,CPMM}, centroid molecular dynamics (CMD)\cite{CMD,HRV,PTM} and ring polymer molecular dynamic (RPMD)\cite{CMQ,MMQ,HMM,RCM,WCR},   make use of the imaginary-time path integral formalism and exploit the exact equilibrium mapping between a quantum-mechanical particle and a classical ring polymer. Thus various of techniques of MD simulations can be directly implemented in PIMD simulations\cite{PTM,CPMM}.

One of the major problems in the PIMD simulations is to integrate the ring polymer Hamiltonian dynamics in the Cartesian coordinate. As the integrated systems are chaotic, it's not possible to obtain accurate trajectories for more than a short time interval. Rather, we expect to generate the trajectories that satisfy correct statistical properties, such as near-conservation of the Hamiltonian and preservation of the wedge product. In practice, these properties can be achieved by symplectic and time-reversible  integrators\cite{IRS}.

Another problem in real simulations is that the harmonic oscillations of the beads and the fast bonded force restrict the time step, whereas the evaluations of slow non-bonded forces accounts for most of the computational time. A wise way is to integrate different components of force field using different time steps, termed the multiple time stepping scheme (MTS)\cite{GHWS,HFB,LRS,HLW}.  The numerical integrators with variable time steps have facilitated the inclusions of full electrostatic forces and Lennard-Jones interactions. In addition, since the harmonic oscillations among beads can be solved exactly by normal mode analysis, the MTS scheme (also termed as trigonometric methods) preserves the Hamiltonian much better than traditional Verlet/leapfrog method.

However, the MTS scheme still suffers from the numerical resonances when the frequency of slow force impulse coincides with a natural frequency of the system, which may lead to an accuracy reduction\cite{IRS,MIS}. This instability may be more severe in ring polymer Hamiltonian dynamics, as the frequency of non-bond force is comparable to that of the harmonic interactions within beads. To ameliorate this problem, it's proposed to treat small covalent molecules as a set of rigid bodies\cite{WCR,ATB}, which not only significantly reduces the degrees of freedom required to represent the system, but also removes the intramolecular vibrations. The price to pay is to impose several holomonic constraints on the Hamiltonian dynamics, which needs to be preserved in the numerical integrations. In the Cartesian coordinates, this problem can be solved efficiently by SHAKE\cite{RCB} and its velocity version RATTLE\cite{AR}, and these algorithms have been throughly analyzed in \cite{LRS,LSS,RS}. Several methods based on SHAKE are proposed to achieve better convergence\cite{GP,BLS}. For the holonomic constraints in more complicated geometries, the corresponding integrators are discussed in \cite{WCR,DLMS}.

In this paper, we focus on numerical integrations of constrained ring-polymer Hamiltonian dynamics in the Cartesian coordinate, where only holonomic constraints are considered. Since the system consists of a fictitious ring polymer connected by stiff harmonic springs, the traditional SHAKE and RATTLE will not conserve the Hamiltonian well unless the time step is very small\cite{CPMM}. We propose a trigonometric method based on the splitting of operators. This method integrates the ring polymer Hamiltonian dynamics under normal mode representation and preserves the constraints by solving an algebraic equation iteratively. It shows that this integrator allows variable time steps and the use of mollified forces, which origin from the mollified impulse method\cite{IRS,HLW,GSS,SM}. We also analyze its near-conservation of Hamiltonian in the framework of modulated Fourier expansion\cite{HLW,CHL1,CHL2,CJLL}, with its accuracy presented by numerical tests.

The rest of this paper is organized as follows. Section 2 begins by briefly reviewing the theory of PIMD. Section 3 presents the formulation of the trigonometric method, and the preservation of the symplectic structure and time-reversibility is also discussed. Section 4 presents the numerical results, with a conclusion drawn in Section 5. 

\section{Background}
In this section we briefly review the methodology of PIMD\cite{HMM}. The Hamiltonian of a quantum system with $N$ degrees of freedom is 
\begin{equation}
\hat{H}\left(\bm{p},\bm{x}\right)=\sum_{i=1}^{N}\frac{p_{i}^{2}}{2m}+V\left(x_{1},...,x_{N}\right).
\end{equation}

Denote the inverse thermal energy by $\beta=1/k_{B}T$ and the quantum canonical partition function is expressed as
\begin{equation}\label{QCPF}
Z=tr\left[e^{-\beta \hat{H}}\right].
\end{equation}

By exploiting the classical isomorphism between path integral representation of quantum mechanical partition function and classical partition function of a ring polymer\cite{ST}, Eq.\eqref{QCPF} can be approximated by the discrete path integral representation as
\begin{equation}
Z=Z_{P}+\mathcal{O}\left(1/P^{2}\right),
\end{equation}
 with
\begin{equation}
Z_{P}=\frac{1}{\left(2\pi \hbar\right)^{NP}} \int d \bm{p} \int d \bm{x} e^{-\beta_{P}H_{P}\left( \bm{p}, \bm{x}\right)},
\end{equation}
where  $P$ is the Trotter number and $\beta_{P}=\beta /P$.
\begin{equation}\label{ring_polymer_Hamiltonian}
H_{P}\left(\bm{p},\bm{x}\right)=\sum_{j=1}^{N}\sum_{k=1}^{P}\left[\frac{\left(p_{j}^{\left(k\right)}\right)^{2}}{2m_{j}}+\frac{m_{j}}{2\beta_{P}^{2} \hbar^{2}}\left(x_{j}^{\left(k\right)}-x_{j}^{\left(k-1\right)}\right)^{2}\right]+\sum_{k=1}^{P}V\left(x_{1}^{\left(k\right)},...,x_{N}^{\left(k\right)}\right),
\end{equation}
subject to the cyclic boundary condition $x_{j}^{(0)}=x_{j}^{(P)}$\cite{DMPM}. We denote by the subscript $j$ and superscript $k$ of $x_{j}^{(k)}$ the $j$th degree of freedom of the $k$th replica. 

The implementation of PIMD (including CMD and RPMD) involves obtaining trajectories $x_{j}^{\left(k\right)}\left(t\right)$ from the time evolution of the ring polymer Hamiltonian \eqref{ring_polymer_Hamiltonian}
\begin{equation}
\dot{\bm{p}}=-\frac{\partial H_{P}\left(\bm{p},\bm{x}\right)}{\partial \bm{x}},~~\dot{\bm{x}}=\frac{\partial H_{P}\left(\bm{p},\bm{x}\right)}{\partial \bm{p}}.
\end{equation}
The corresponding equations of motion (EOMs) are
\begin{equation}\label{EOMs}
\begin{split}
&\dot{p}_{j}^{\left(k\right)}=-\frac{m_{j}}{\beta_{P}^{2} \hbar^{2}}\left[2x_{j}^{\left(k\right)}-x_{j}^{\left(k-1\right)}-x_{j}^{\left(k+1\right)}\right]-\frac{\partial V\left(x_{1}^{\left(k\right)},\cdots,x_{N}^{\left(k\right)}\right)}{\partial x_{j}^{\left(k\right)}}, \\
&\dot{x}_{j}^{\left(k\right)}=\frac{p_{j}^{\left(k\right)}}{m_{j}}. \\
\end{split}
\end{equation}
Eq.\eqref{EOMs} are equivalent to second order ODEs
\begin{equation}
\ddot{x}_{j}^{\left(k\right)}=-\frac{1}{\beta_{P}^{2} \hbar^{2}}\left[2x_{j}^{\left(k\right)}-x_{j}^{\left(k-1\right)}-x_{j}^{\left(k+1\right)}\right]-\frac{1}{m_{j}}\frac{\partial V\left(x_{1}^{\left(k\right)},\cdots,x_{N}^{\left(k\right)}\right)}{\partial x_{j}^{\left(k\right)}}.
\end{equation}    

$V\left(\bm{x}\right)$ can be chosen as either empirical atomic potentials or the Kohn-Sham functional\cite{DMPM,LLS}. We only consider atomic potentials, which are typically given by
\begin{equation}
V=V^{\textup{bond}}+V^{\textup{Lennard-Jones}}+V^{\textup{electrostatic}},
\end{equation}
where bonded force corresponds to highly oscillatory motions, whereas nonbonded forces have mostly low-frequency motions\cite{IRS}.

Since the high-frequency intramolecular forces may give rise to resonances in the Hamiltonian dynamics \eqref{EOMs}, in many applications we can fixate the bond lengths and angles, thereby removing the fast bonded forces. The rigid bond models significantly simplify the evaluations of $V\left(\bm{x}\right)$, at the cost of imposing several holonomic constraints $g\left(\bm{x}\right)=0$ on the Hamiltonian systems. In the next section we will give the formulation of the trigonometric integrators and illustrate it with a simple example.

\section{Ring polymer time evolution}
We consider the holonomic constraints
\begin{equation}\label{constraints}
g\left(\bm{x}^{(k)}\right)=0,
\end{equation}
$g :  \mathbb{R}^{N} \rightarrow \mathbb{R}^{M}$ and hidden velocity constraints
\begin{equation}
f\left(\bm{p}^{(k)},\bm{x}^{(k)}\right)=G\left(\bm{x}^{(k)}\right)\nabla_{\bm{p}^{(k)}}H\left(\bm{p}^{(k)},\bm{x}^{(k)}\right)=0,
\end{equation}    
where $G\left(\bm{x}^{(k)}\right)=g_{\bm{x}^{(k)}}\left(\bm{x}^{(k)}\right) \in \mathbb{R}^{M \times N}$ and $\left(\bm{p}^{(k)},\bm{x}^{(k)}\right)$ are the momentum and position of $k$th replicas, respectively. 

Eq.\eqref{EOMs} and \eqref{constraints} define a Hamiltonian vector field on the $(2NP-2MP)$ manifold
\begin{equation}\label{constrained_Hamiltonian}
\mathcal{M}=\left\{\left(\bm{p},\bm{x}\right):g\left(\bm{x}^{(k)}\right)=0, G\left(\bm{x}^{(k)}\right)\nabla_{\bm{p}^{(k)}}H\left(\bm{p}^{(k)},\bm{x}^{(k)}\right)=0, k=1,\cdots,P\right\}.
\end{equation}
The symplectic structure on $\mathcal{M}$ is defined by the differential 2-form restricted on $\mathcal{M}$\cite{LSS}.

To derive the equations of motion for the constrained Hamiltonian \eqref{constrained_Hamiltonian}, it is proposed to add Lagrangian multipliers that grow large when system deviates from the locus of the constraints\cite{RCB,AR}.  The modified Hamiltonian $H_{\mathcal{M}}$ is expressed as
\begin{equation}\label{constrained_RPH}
H_{\mathcal{M}}\left(\bm{p}, \bm{x}\right)=H_{P}\left(\bm{p}, \bm{x}\right)+\sum_{k=1}^{P}\frac{1}{2\epsilon} g\left(\bm{x}^{(k)}\right)^{\tau}g\left(\bm{x}^{(k)}\right),
\end{equation}
with $0<\epsilon\ll 1$ and  $\bm{x}^{(k)}=\left(x_{1}^{(k)},x_{2}^{(k)},\cdots,x_{N}^{(k)}\right)$. The corresponding EOMs in the Cartesian coordinate are
\begin{equation}\label{EOMs_constraints}
\begin{split}
&\dot{p}_{j}^{\left(k\right)}=-\frac{m_{j}}{\beta_{P}^{2} \hbar^{2}}\left[2x_{j}^{\left(k\right)}-x_{j}^{\left(k-1\right)}-x_{j}^{\left(k+1\right)}\right]-\frac{\partial V}{\partial x_{j}^{\left(k\right)}}-g_{\bm{x}}\left(\bm{x}^{(k)}\right)^{\tau}\cdot \bm{\Lambda^{\left(k\right)}}, \\
&\dot{x}_{j}^{\left(k\right)}=\frac{p_{j}^{\left(k\right)}}{m_{j}}, \\
&\bm{\Lambda}^{(k)}= \frac{1}{\epsilon}g\left(\bm{x}^{(k)}\right).
\end{split}
\end{equation}
When $\epsilon \to 0$, it gives rise to the constrained Hamiltonian system\cite{RS}.

Traditionally, the EOMs \eqref{EOMs_constraints} are integrated by Verlet/leapfrog algorithm and the constraints Eq.\eqref{constraints} are solved successively by SHAKE or RATTLE algorithm\cite{RCB,AR}. The iteration will not end until all the constraints are satisfied. This approach is very efficient for large molecules, but also suffers from non-convergence when the distortion is large enough. For the molecules with simple topology, it's better to tackle all the constraints simultaneously, such as the Matrix Inverted Linearized Constraints (MILC) algorithm\cite{BLS}.

However, the RATTLE algorithm will not conserve the Hamiltonian \eqref{constrained_Hamiltonian} very well unless the time step is sufficiently small. Therefore, we need to make a modification on the RATTLE algorithm under the normal mode representation of Eq.\eqref{EOMs_constraints}, which greatly facilitates the integration of unconstrained problem \eqref{EOMs}.  It shows that the normal mode theory is also applicable for the constrained problem \eqref{EOMs_constraints}. 

\subsection{Normal mode representation}
It begins by solving the linear part of Eq.\eqref{EOMs_constraints} 
\begin{equation}\label{normal_mode}
\begin{split}
&\dot{p}_{j}^{\left(k\right)}=-\frac{m_{j}}{\beta_{P}^{2} \hbar^{2}}\left[2x_{j}^{\left(k\right)}-x_{j}^{\left(k-1\right)}-x_{j}^{\left(k+1\right)}\right], \\
&\dot{x}_{j}^{\left(k\right)}=\frac{p_{j}^{\left(k\right)}}{m_{j}},
\end{split}
\end{equation}

For a fixed degree of freedom $j$, we denote $\bm{p}_{j}=\left(p_{j}^{(1)},\cdots, p_{j}^{(P)}\right)^{\tau}$, $\bm{x}_{j}=\left(x_{j}^{(1)},\cdots, x_{j}^{(P)}\right)$. Thus Eq.\eqref{normal_mode} is rewritten in matrix formalism
\begin{equation}\label{linear_matrix_eq}
\begin{pmatrix} \dot{\bm{p}_{j}}\\ \dot{\bm{x}_{j}}\end{pmatrix}= \begin{pmatrix} 0 & K_{1}\\ K_{2} & 0 \end{pmatrix} \begin{pmatrix} \bm{p}_{j}\\ \bm{x}_{j} \end{pmatrix},
\end{equation}
with $K_{2}=\displaystyle{\frac{1}{m_{j}}} I$ and
\begin{equation}
K_{1}=\displaystyle{\frac{m_{j}}{\beta_{P}^{2} \hbar^{2}}} \begin{pmatrix} -2 & 1  & \cdots & 1\\ 1 & -2 & \cdots & 1 \\ \vdots & \ddots & & \vdots\\ 1 & \cdots & 1 & -2 \end{pmatrix}.
\end{equation}

Since $K_{1}$ can be diagonalized by trigonometric basis 
\begin{equation}
\left\{1, \sin\frac{\pi}{P}, \cos\frac{\pi}{P}, \cdots, \sin\frac{(P-1)\pi}{P}, \cos\frac{(P-1)\pi}{P}\right\},
\end{equation}
there exists a unitary matrix $U$ such that 
\begin{equation}
U^{\tau}K_{1}U=D, 
\end{equation}
where the eigenvalues of diagonal matrix $D$ is $\displaystyle {\left\{0, -4\alpha^{2} \sin^{2}\frac{\pi}{P}, \cdots, -4\alpha^{2} \sin^{2}\frac{\left(n-1\right)\pi}{P}\right\}}$ and  $\alpha=P/\beta \hbar$\cite{CPMM}. 

By taking $\bm{\tilde{x}_{j}}=U\bm{x}_{j}$ and $\bm{\tilde{p}_{j}}=U\bm{p}_{j}$ (termed the normal mode representation), we arrive at the exact solution of Eq.\eqref{normal_mode},
\begin{eqnarray}\label{normal_mode_representation}
\begin{split}
&\begin{pmatrix} x_{j}^{(1)}\left(t\right)\\x_{j}^{(2)}\left(t\right) \\ \vdots \\ x_{j}^{(n)}\left(t\right) \end{pmatrix}= \hat{A}\left(t\right) \begin{pmatrix} x_{j}^{(1)}\left(0\right)\\x_{j}^{(2)}\left(0\right) \\ \vdots \\ x_{j}^{(n)}\left(0\right) \end{pmatrix}+\frac{1}{m_{j}}
\hat{B}\left(t\right) \begin{pmatrix} p_{j}^{(1)}\left(0\right)\\p_{j}^{(2)}\left(0\right) \\ \vdots \\ p_{j}^{(n)}\left(0\right)\end{pmatrix},\\
&\begin{pmatrix} p_{j}^{(1)}\left(t\right)\\p_{j}^{(2)}\left(t\right) \\ \vdots \\ p_{j}^{(n)}\left(t\right) \end{pmatrix}= m_{j}\hat{C}\left(t\right) \begin{pmatrix} x_{j}^{(1)}\left(0\right)\\x_{j}^{(2)}\left(0\right) \\ \vdots \\ x_{j}^{(n)}\left(0\right) \end{pmatrix}+
\hat{A}\left(t\right) \begin{pmatrix} p_{j}^{(1)}\left(0\right)\\p_{j}^{(2)}\left(0\right) \\ \vdots \\ p_{j}^{(n)}\left(0\right)\end{pmatrix},
\end{split}
\end{eqnarray}
with $\hat{A}\left(t\right)=UA\left(t\right)U^{\tau}$, $\hat{B}\left(t\right)=UB\left(t\right)U^{\tau}$, $\hat{C}\left(t\right)=UC\left(t\right)U^{\tau}$.

$A\left(t\right), B\left(t\right), C\left(t\right)$ are diagonal matrices, defined by
\begin{align*}
&A\left(t\right)=\textup{diag}\left\{1, \cos\left(\omega_{1} t\right), \cdots, \cos\left(\omega_{n-1} t\right)\right\},\\
&B\left(t\right)=\textup{diag}\left\{1, \sin\left(\omega_{1} t\right)/\omega_{1} , \cdots,  \sin\left(\omega_{n-1} t\right)/\omega_{n-1}\right\},\\
&C\left(t\right)=\textup{diag}\left\{0, -\omega_{1} \sin\left(\omega_{1} t\right), \cdots, -\omega_{n-1} \sin\left(\omega_{n-1} t\right)\right\},
\end{align*}
where $\displaystyle{\omega_{k}=2\alpha \sin\frac{k\pi}{P}}\left(k=1,2,\cdots ,P-1\right)$. 

For the nonlinear problem Eq.\eqref{EOMs} and constrained problem Eq.\eqref{EOMs_constraints}, we 
can derive the trigonometric integrators from the variation-of-constant formula.

\subsection{The trigonometric integrators for the constrained formulation}
The trigonometric integrators are based on the splitting of the Hamiltonian \eqref{constrained_RPH}  according to their natural frequencies,
\begin{equation}
H_{\mathcal{M}}\left(\bm{p},\bm{x}\right)=H_{0}\left(\bm{p},\bm{x}\right)+V\left(\bm{x}\right)+V_{c}\left(\bm{x}\right),
\end{equation}
where
\begin{align}
\begin{split}
&H_{0}\left(\bm{p},\bm{x}\right)=\sum_{j=1}^{N}\sum_{k=1}^{P}\left[\frac{\left(p_{j}^{\left(k\right)}\right)^{2}}{2m_{j}}+\frac{m_{j}}{2\beta_{P}^{2} \hbar^{2}}\left(x_{j}^{\left(k\right)}-x_{j}^{\left(k-1\right)}\right)^{2}\right],\\
&V\left(\bm{x}\right)=\sum_{k=1}^{P}V\left(x_{1}^{\left(k\right)},\cdots,x_{N}^{\left(k\right)}\right),\\
&V_{c}\left(\bm{x}\right)=\sum_{k=1}^{P}\frac{1}{2\epsilon} g\left(x_{1}^{\left(k\right)},\cdots,x_{N}^{\left(k\right)}\right)^{\tau}g\left(x_{1}^{\left(k\right)},\cdots,x_{N}^{\left(k\right)}\right).
\end{split}
\end{align}
We assume that the eigenfrequency of $\nabla^{2}_{\bm{x}}V\left(\bm{x}\right)$ is much smaller than that of $\nabla^{2}_{\bm{x}}H_{0}\left(\bm{p},\bm{x}\right)$.
 
Denote by $L$ the Liouvillian associated with the Hamiltonian $H\left(\bm{p},\bm{x}\right)$ and the operator propagator $\varphi_{h}^{H}=e^{-h L}$, with $h$ a small time step. Owing to the Trotter formula, $\varphi_{h}^{H}$ can be approximated by the symmetric composition of subflows,
\begin{equation}\label{composition_of_flows}
\varphi_{h}^{H_{c}} \approx \varphi_{h/2}^{V_{c}}\circ \varphi_{h/2}^{V}  \circ \varphi_{h}^{H_{0}} \circ \varphi_{h/2}^{V} \circ \varphi_{h/2}^{V_{c}}.
\end{equation}
$\varphi_{h}^{H_{0}}$ is determined by Eq.\eqref{normal_mode}. In this case, it can be solved exactly.

Thus we suggest the following integration scheme, with $\bm{p}^{(k)}_{n}$ and $\bm{x}^{(k)}_{n}$ the abbreviations of $\left(p_{1}^{(k)}, \cdots, p_{N}^{(k)}\right)$ and $\left(x_{1}^{(k)}, \cdots, x_{N}^{(k)}\right)$ at $t=t_{n}$, respectively.

\begin{algorithm}\label{scheme_1}The trigonometric integrator with constant time step

Step 1.
\begin{align*}
&\bm{\bar{p}}_{n}^{(k)}=\bm{p}_{n}^{(k)}-\frac{h}{2} \nabla_{\bm{x}^{(k)}} V\left(\bm{x}_{n}^{(k)}\right)-\frac{h}{2}g_{\bm{x}^{(k)}}\left(\bm{x}_{n}^{(k)}\right)^{\tau}\cdot \Lambda_{c}^{\left(k\right)},\\
&\bm{\bar{x}}_{n}^{(k)}=\bm{x}_{n}^{(k)},
\end{align*}

Step 2.
\begin{align*}
\begin{pmatrix}\bm{\bar{p}}^{(k)}_{n+1}\\\bm{\bar{x}}^{(k)}_{n+1}\end{pmatrix}=\varphi_{h}^{H_{0}}\begin{pmatrix}\bm{\bar{p}}^{(k)}_{P}\\\bm{\bar{x}}^{(k)}_{P}\end{pmatrix}
\end{align*}

Step 3.
\begin{align*}
&\bm{p}_{n+1}^{(k)}=\bm{\bar{p}}_{n+1}^{(k)}-\frac{h}{2} \nabla_{\bm{x}^{(k)}} V\left(\bm{\bar{x}}_{n+1}^{(k)}\right)-\frac{h}{2}g_{\bm{x}^{(k)}}\left(\bm{\bar{x}}_{n+1}^{(k)}\right)^{\tau}\cdot \Lambda_{cv}^{\left(k\right)},\\
&\bm{x}_{n+1}^{(k)}=\bm{\bar{x}}_{n+1}^{k},
\end{align*}
where $\Lambda_{c}^{(k)}$ and $\Lambda_{cv}^{(k)}$ are chosen to satisfy
\begin{align*}
g\left(\bm{x}^{(k)}_{n+1}\right)=0,
\end{align*}
and
\begin{align*}
f\left(\bm{p}^{(k)}_{n+1},\bm{x}^{(k)}_{n+1}\right)=0,
\end{align*}
respectively.
\end{algorithm} 

Since the second step is solved exactly, each flow mapping is symplectic and their composition is also symplectic\cite{LSS}. Moreover, the scheme \ref{scheme_1} is time-reversible due to the symmetric structure. 

Although the trigonometric integrator allows a longer time step for the unconstrained problem \eqref{EOMs}, its stability may be contaminated when the frequency of slow force impulse coincides with the natural frequency of the system, leading to an oscillation in the positions with an increasing amplitude (known as numerical resonance)\cite{IRS,MIS}. Intuitively speaking, it is because the slow force is only evaluated at the end of each time step, but doesn't enter into the oscillations. A similar problem will occur when integrating the constrained Hamiltonian dynamics using the scheme \ref{scheme_1}.

In the previous studies, there exists various of methods to overcome this stability barrier, including mollified impulse method\cite{HLW,LRS,GSS}, adiabatic separation\cite{HRV} and normal mode theories \cite{CMQ}. It's also known that the numerical resonance is less severe in the Langevin dynamics and Nos\'{e}-Hoover thermostatting\cite{CPMM,OK}. In this work, we only discuss the application of mollified slow force in the constrained dynamics.  

The mollified impulse method is given by replacing the slow potential $V\left(\bm{x}^{(k)}\right)$ with a mollified potential $\tilde{V}_{nb}\left(\bm{x}^{(k)}\right)=V\left(\mathcal{A}\left(\bm{x}^{(k)}\right)\right)$, so that the force is evaluated at an averaged position $\mathcal{A}\bm{x}^{(k)}$, instead of several isolated points. The choice of averaging operator $\mathcal{A}: \mathbb{R}^{P\times N} \rightarrow \mathbb{R}^{P \times N}$ can be founded in\cite{IRS,HLW,GSS,SM}.  

For instance, one can solve the auxiliary initial value position
\begin{equation}
\ddot{\tilde{x}}_{j}^{\left(k\right)}=-\frac{1}{\beta_{P}^{2} \hbar^{2}}\left[2\tilde{x}_{j}^{\left(k\right)}-\tilde{x}_{j}^{\left(k-1\right)}-\tilde{x}_{j}^{\left(k+1\right)}\right]~~\left(k=1,\cdots, P\right) 
\end{equation} 
with $\tilde{x}_{j}^{(k)}\left(0\right)=x_{j}^{(k)}$, $\dot{x}_{j}^{(k)}\left(0\right)=0$. Then $\mathcal{A}$ is defined by 
\begin{equation}
\mathcal{A}\left(x_{j}^{(k)}\right)=\frac{1}{h}\int_{0}^{h} \tilde{x}_{j}^{(k)}\left(t\right)dt.
\end{equation}

Due to the normal mode representation \eqref{normal_mode_representation}, the averaging operator has an explicit form
\begin{equation}
\mathcal{A}\left(\bm{x}_{j}\right)=UD\left(h\right)U^{\tau} \bm{x}_{j},
\end{equation}
with  $D\left(h\right)=\textup{diag}\left\{1, sin\left(\omega_{1} h\right)/\omega_{1}h , \cdots,  sin\left(\omega_{n-1} h\right)/\omega_{n-1}h\right\}$. The Jacobian matrix of $\mathcal{A}$ is expressed as 
\begin{equation}
\mathcal{A}_{\bm{x}}=UD\left(h\right)U^{\tau}.
\end{equation}

Now we can make a slight modification on the splitting of $\varphi_{h}^{H_{c}}$, 
\begin{equation}
\varphi_{h}^{H_{c}} \approx \varphi_{h/2}^{V_{c}}\circ \varphi_{h/2}^{\tilde{V}} \circ \varphi_{h}^{H_{0}} \circ \varphi_{h/2}^{\tilde{V}} \circ \varphi_{h/2}^{V_{c}},
\end{equation}
yielding the following scheme.

\begin{algorithm}\label{scheme_2}The trigonometric integrator with mollified forces 

Step 1.
\begin{align*}
&\bm{\bar{p}}_{n}^{(k)}=\bm{p}_{n}^{(k)}-\frac{h}{2} \left(\mathcal{A}_{\bm{x}}\right)^{\tau}\nabla_{\bm{x}^{(k)}} V\left(\mathcal{A}\bm{x}_{n}^{(k)}\right)-\frac{h}{2}g_{\bm{x}^{(k)}}\left(\bm{x}_{n}^{(k)}\right)^{\tau}\cdot \Lambda_{c}^{\left(k\right)},\\
&\bm{\bar{x}}_{n}^{(k)}=\bm{x}_{n}^{(k)},
\end{align*}

Step 2.
\begin{align*}
\begin{pmatrix}\bm{\bar{p}}^{(k)}_{n+1}\\\bm{\bar{x}}^{(k)}_{n+1}\end{pmatrix}=\varphi_{h}^{H_{0}}\begin{pmatrix}\bm{\bar{p}}^{(k)}_{n}\\\bm{\bar{x}}^{(k)}_{n}\end{pmatrix}
\end{align*}

Step 3.
\begin{align*}
&\bm{p}_{n+1}^{(k)}=\bm{\bar{p}}_{n+1}^{(k)}-\frac{h}{2} \left(\mathcal{A}_{\bm{x}}\right)^{\tau}\nabla_{\bm{x}^{(k)}} V\left(\mathcal{A}\bm{\bar{x}}_{n+1}^{(k)}\right)-\frac{h}{2}g_{\bm{x}^{(k)}}\left(\bm{\bar{x}}_{n+1}^{(k)}\right)^{\tau}\cdot \Lambda_{cv}^{\left(k\right)},\\
&\bm{x}_{n+1}^{(k)}=\bm{\bar{x}}_{n+1}^{(k)},
\end{align*}
with $\Lambda_{c}^{(k)}$ and $\Lambda_{cv}^{(k)}$ chosen to satisfy
$g\left(\bm{x}^{(k)}_{n+1}\right)=0$ and $f\left(\bm{p}^{(k)}_{n+1},\bm{x}^{(k)}_{n+1}\right)=0$.
\end{algorithm} 

The mollified impulse method is the impulse method with a mollified potential, thus it's  also symplectic and time-reversible.

An alternative way to ameliorate the numerical resonance is to use variable time steps, so that the fast part of non-bonded force is integrated using a smaller time step. It is motivated by  artificially splitting the non-bonded force $V$ into two parts corresponding to their frequencies,
\begin{equation}
V=V_{fast}+V_{slow}.
\end{equation}

We choose a smaller time step $\delta h$ that satisfies $\delta h= h/m$, then integrate $V_{fast}$ and $V_{slow}$ with $h$ and $\delta h$, respectively. The Lagrangian multiplier $V_{c}$ should be integrated using a smaller time step, due to its stiffness. In sum, the flow $\varphi_{h}^{H_{c}}$ is split as
\begin{equation}
\varphi_{h}^{H_{c}} \approx \varphi_{h/2}^{V_{slow}} \circ \left[\varphi_{\delta h/2}^{V_{fast}}\circ \varphi_{\delta h/2}^{V_{c}}  \circ \varphi_{\delta h}^{H_{0}}\circ \varphi_{\delta h/2}^{V_{c}} \circ \varphi_{\delta h/2}^{V_{fast}}\right]^{m} \circ \varphi_{h/2}^{V_{slow}}.
\end{equation}

\begin{algorithm}\label{scheme_3}The trigonometric integrator with multiple time steps  

1. outer loop: 
\begin{equation*}
p_{j}^{\left(k\right)}\leftarrow p_{j}^{\left(k\right)}-\frac{h}{2}  \frac{\partial V_{slow} }{\partial x_{j}^{(k)}} \left(x_{1}^{\left(k\right)},\cdots,x_{N}^{\left(k\right)}\right),
\end{equation*}

2. inner loop:

Step.1
\begin{equation*}
\tilde{p}_{j}^{\left(k\right)}\leftarrow p_{j}^{\left(k\right)}-\frac{\delta h}{2}  \frac{\partial V_{fast}}{\partial x_{j}^{(k)}}  \left(x_{1}^{\left(k\right)},\cdots ,x_{N}^{\left(k\right)}\right)-\frac{\delta h}{2} g_{\bm{x}}\left(x_{1}^{\left(k\right)},\cdots ,x_{N}^{\left(k\right)}\right)^{\tau} \cdot \Lambda^{(k)}_{c},
\end{equation*}

Step.2
\begin{equation*}
\begin{pmatrix} \tilde{p}_{j}^{\left(k\right)} \\ x_{j}^{\left(k\right)} \end{pmatrix} \leftarrow \varphi_{\delta h}^{H_{0}} \begin{pmatrix} \tilde{p}_{j}^{\left(k\right)} \\ x_{j}^{\left(k\right)} \end{pmatrix},
\end{equation*}

Step.3
\begin{equation*}
p_{j}^{\left(k\right)}\leftarrow \tilde{p}_{j}^{\left(k\right)}-\frac{\delta h}{2}  \frac{\partial V_{fast}}{\partial x_{j}^{(k)}}  \left(x_{1}^{\left(k\right)},\cdots,x_{N}^{\left(k\right)}\right)-\frac{\delta h}{2} g_{\bm{x}}\left(x_{1}^{\left(k\right)},\cdots ,x_{N}^{\left(k\right)}\right)^{\tau} \cdot \Lambda^{(k)}_{cv},
\end{equation*}

3.  outer loop:
\begin{equation*}
{p}_{j}^{\left(k\right)}\leftarrow p_{j}^{\left(k\right)}-  \frac{\delta h}{2}  \frac{\partial V_{slow} }{\partial x_{j}^{(k)}} \left(x_{1}^{\left(k\right)},\cdots ,x_{N}^{\left(k\right)}\right).
\end{equation*}

$\Lambda_{c}$ and $\Lambda_{cv}$ are chosen to satisfy the constraints
\begin{align*}
&g\left(x_{1}^{(k)},\cdots,x_{N}^{(k)}\right)=0,\\
&f\left(p_{1}^{(k)},\cdots, p_{N}^{(k)},x_{1}^{(k)},\cdots,x_{N}^{(k)}\right)=0.
\end{align*}

\end{algorithm} 

The scheme \ref{scheme_3} is also symplectic and time-reversible. Furthermore, we can replace slow force $\nabla_{\bm{x}^{(k)}} V_{slow}\left(\bm{x}_{n}^{(k)}\right)$ by a mollified force $ \left(\mathcal{A}_{\bm{x}}\right)^{\tau}\nabla_{\bm{x}^{(k)}} V_{slow}\left(\mathcal{A}\bm{x}_{n}^{(k)}\right)$ to remove the distabilizing components of the slow force.

Finally, we give the explicit formulae for the above schemes. For the $j$th degree of freedom, $\bm{p}_{j}$ and $\bm{x}_{j}$ at $t=t_{n}$ are denoted by $\bm{p}_{j,n}$ and $\bm{x}_{j,n}$, respectively. Combining Eq.\eqref{normal_mode_representation} and the variation-of-constants formula, we can write the schemes \ref{scheme_1} and \ref{scheme_2} in the two-step form 

\begin{align}\label{two-step.form}
\begin{split}
\bm{x}_{j,n}=&\hat{A}\left(h\right) \bm{x}_{j,n}+\frac{1}{m_{j}} \hat{B}\left(h\right) \bm{p}_{j,n}-\frac{1}{2m_{j}}h^{2}\hat{B}\left(h\right) \bm{u}_{j,n}-\frac{1}{2m_{j}}h^{2}\hat{B}\left(h\right)\phi\left(h\right) \bm{v}_{j,n}^{\tau}\bm{\Lambda}^{c}_{j,n}\\
\bm{p}_{j,n}=&m_{j}\hat{C}\left(h\right) \bm{x}_{j,n}+\hat{A}\left(h\right) \bm{p}_{j,n}-\frac{1}{2}h \hat{A}\left(h\right) \phi\left(h\right)\bm{u}_{j,n}-\frac{1}{2}h \phi\left(h\right) \bm{u}_{j,n+1}\\
&-\frac{1}{2} h\hat{A}\left(h\right) \bm{v}_{j,n}^{\tau} \bm{\Lambda}_{n}^{c}- \frac{1}{2}h  \bm{v}_{j,n+1}^{\tau} \bm{\Lambda}_{j,n+1}^{cv}
\end{split}
\end{align}
with 
\begin{align*}
&\bm{u}_{j,n}=\left(\nabla_{{x}_{j}^{(1)}}V\left(\phi\left(h\Omega\right)\bm{x}_{1,n} \cdots  \phi\left(h\Omega\right)\bm{x}_{N,n}\right), \cdots, \nabla_{{x}_{j}^{(P)}}V\left(\phi\left(h\Omega\right)\bm{x}_{1,n} \cdots  \phi\left(h\Omega\right)\bm{x}_{N,n}\right)\right)^{\tau},\\ 
&\bm{v}_{j,n}=\left(g_{{x}_{j}^{(1)}}\left(\bm{x}_{1,n} \cdots \bm{x}_{N,n}\right), \cdots, g_{{x}_{j}^{(P)}}\left(\bm{x}_{1,n} \cdots \bm{x}_{N,n}\right)\right)^{\tau}.
\end{align*}
and in the scheme \ref{scheme_2}, $\phi\left(h\right)=UD\left(h\right)U^{\tau}$, while in the scheme \ref{scheme_1} $\phi\left(h\right)$ is replaced by an identity matrix. The explicit formula of the scheme \ref{scheme_3} can be derived in a similar way.

It remains to choose $\Lambda^{(k)}_{c}$ and $\Lambda^{(k)}_{cv}$ so that $\bm{p}^{(k)}_{n+1}$ and 
$\bm{x}^{(k)}_{n+1}$ satisfy the constraints $g\left(\bm{x}_{n+1}^{(k)}\right)$ and $f\left(\bm{p}_{n+1}^{(k)}, \bm{x}_{n+1}^{(k)}\right)=0$. In practice, they can be obtained by solving nonlinear equations using iterative Newton method, in the spirit of SHAKE and RATTLE algorithms. The initial guess is made by taking $\Lambda^{(k)}_{c}=\bm{0}$ and $\Lambda^{(k)}_{cv}=\bm{0}$ for all $k$. Since the iteration becomes a little more complicated under normal mode representation, we put the detailed discussions later.

\subsection{A simple example}
In this part, we will illustrate how to solve $\Lambda^{(k)}_{c}$ and $\Lambda^{(k)}_{cv}$. We consider the system composed of water molecules with extended simple charge potential (SPC/E model) \cite{BPVG}. This example is motivated by the simulations of the quantum diffusion of water molecules using the ring-polymer molecular dynamics\cite{MMQ}. 

Since each water molecule is composed of three atoms with a simple ring topology, we can fixate the bond lengths of O-H bond and H-H bond, yielding three constraints for $kth$ replica
\begin{equation}\label{bond_constraints}
\left|\bm{r}^{(k)}_{i}-\bm{r}^{(k)}_{j}\right|^{2}=l_{ij}^{2},
\end{equation}
with $\bm{r}^{(k)}_{i}=\left(x_{3i-2}^{(k)}, x_{3i-1}^{(k)}, x_{3i}^{(k)}\right)$ is a line vector that presents the position of the $i$th atom of the $k$th replica, and $l_{ij}$ the bond length. 

By differentiating Eq.\eqref{bond_constraints}, we get the velocity constraints,
\begin{equation}\label{water_velocity_constraints}
\left(\dot{\bm{r}}_{i}^{(k)}-\dot{\bm{r}}_{j}^{(k)}\right) \cdot \left(\bm{r}_{i}^{(k)}-\bm{r}_{j}^{(k)}\right)=0,
\end{equation}
with $\bm{\dot{r}}^{(k)}_{i}=\left(p_{3i-2}^{(k)}/m_{3i-2}, p_{3i-1}^{(k)}/m_{3i-1}, p_{3i}^{(k)}/m_{3i}\right)$.

Thus, the Lagrangian multiplier $g\left(\bm{x}^{(k)}\right)$ is expressed as
\begin{equation}\label{constraints_water}
g\left(\bm{x}^{(k)}\right)=\frac{1}{2}\sum_{i}\sum_{i \rightarrow p}\lambda^{(k)}_{ip}\left(\left|\bm{r}^{(k)}_{i}-\bm{r}^{(k)}_{j}\right|^{2}-l^{2}_{ij}\right),
\end{equation} 
where the summation is over all sites, indexed by $j$, connected to site $i$. $\Lambda_{c}^{(k)}=\left\{\lambda_{ij}^{(k)}\right\}$ are time-dependent  Lagrangian multipliers, which can be solved by iteration. 

Now we denote the positions of the oxygen atoms and two hydrogen atoms by $\bm{r}_{1}^{(k)}$,  $\bm{r}_{2}^{(k)}$ and $\bm{r}_{3}^{(k)}$. Then pick up a O-H bond constraint   
\begin{equation}\label{eq.water_constraints}
\left|\bm{r}_{12}^{(k)}\right|^{2}:=\left|\bm{r}_{1}^{(k)}-\bm{r}_{2}^{(k)}\right|^{2}=l_{12}^{2} ~~\left(k=1,...,P\right),
\end{equation}
and $\bm{r}_{12}^{(k)}=\left(r_{12,1}^{(1)}, r_{12,2}^{(2)}, r_{12,3}^{(3)}\right)$ satisfy 
\begin{eqnarray}\label{eq.37}
\begin{split}
\begin{pmatrix} r_{12,s}^{(1)}\left(t_{n+1}\right)\\r_{12,s}^{(2)}\left(t_{n+1}\right) \\ \vdots \\ r_{12,s}^{(P)}\left(t_{n+1}\right) \end{pmatrix}= &\begin{pmatrix} \tilde{r}_{12,s}^{(1)}\left(t_{n+1}\right)\\\tilde{r}_{12,s}^{(2)}\left(t_{n+1}\right) \\ \vdots \\ \tilde{r}_{12,s}^{(P)} \left(t_{n+1}\right) \end{pmatrix}+
\frac{h}{m_{12}}\hat{B}\left(h\right) \begin{pmatrix} \lambda_{12,s}^{(1)} r_{12,s}^{(1)}\left(t_{n}\right)\\\lambda_{12,s}^{(2)}  r_{12,s}^{(2)}\left(t_{n}\right) \\ \vdots \\ \lambda_{12,s}^{(P)}  r_{12,s}^{(	P)}\left(t_{n}\right)\end{pmatrix}\\
&-\frac{h}{m_{1}} \hat{B}\left(h\right)\begin{pmatrix} \lambda_{31}^{(1)} r_{31,s}^{(1)}\left(t_{n}\right)\\\lambda_{31}^{(2)}  r_{31,s}^{(2)}\left(t_{n}\right) \\ \vdots \\ \lambda_{31}^{(P)}  r_{31,s}^{(P)}\left(t_{n}\right)\end{pmatrix}-\frac{h}{m_{2}} \hat{B}\left(h\right)\begin{pmatrix} \lambda_{23}^{(1)} r_{23,s}^{(1)}\left(t_{n}\right)\\\lambda_{23}^{(2)}  r_{23,s}^{(2)}\left(t_{n}\right) \\ \vdots \\ \lambda_{23}^{(P)}  r_{23,s}^{(P)}\left(t_{n}\right)\end{pmatrix},
\end{split}
\end{eqnarray} 
with $s=1,2,3$, $m_{12}=\left(m_{1}m_{2}\right)/\left(m_{1}+m_{2}\right)$. $\tilde{r}_{ij,s}^{(k)}$ is the initial guess of $r_{ij,s}^{(k)}$ by putting $\lambda_{ij}^{(k)}=0$.

Combining Eq.\eqref{eq.37} with Eq.\eqref{eq.water_constraints} and omit the second-order term with respect to $\lambda_{ij}^{(k)}$, we arrive at the system of equations,
\begin{equation}
\begin{split}
\begin{pmatrix} l_{12}^{2}\\l_{12}^{2} \\ \vdots \\ l_{12}^{2} \end{pmatrix}&= \sum_{s=1}^{3} \frac{2h}{m_{12}} \begin{pmatrix} \tilde{r}_{12,s}^{(1)} & & & \\ & \tilde{r}_{12,s}^{(2)}& &\\& & \ddots &\\ & & & \tilde{r}_{12,s}^{(P)} \end{pmatrix}\hat{B}\left(h\right)\begin{pmatrix} r_{12,s}^{(1)} & & & \\ & r_{12,s}^{(2)}& &\\& & \ddots &\\ & & & r_{12,s}^{(P)} \end{pmatrix}\begin{pmatrix}  \lambda_{12,s}^{(1)}\\  \lambda_{12,s}^{(2)} \\ \vdots \\   \lambda_{12,s}^{(P)}\end{pmatrix}\\
&-\sum_{s=1}^{3}\frac{h}{m_{1}}\begin{pmatrix} \tilde{r}_{12,s}^{(1)} & & & \\ & \tilde{r}_{12,s}^{(2)}& &\\& & \ddots &\\ & & & \tilde{r}_{12,s}^{(P)} \end{pmatrix}\hat{B}\left(h\right)\begin{pmatrix} r_{31,s}^{(1)} & & & \\ & r_{31,s}^{(2)}& &\\& & \ddots &\\ & & & r_{31,s}^{(P)} \end{pmatrix}\begin{pmatrix} \lambda_{31,s}^{(1)}\\\lambda_{31,s}^{(2)}  \\ \vdots \\ \lambda_{31,s}^{(P)}  \end{pmatrix}\\
&-\sum_{s=1}^{3}\frac{h}{m_{2}} \begin{pmatrix}\tilde{r}_{12,s}^{(1)} & & & \\ & \tilde{r}_{12,s}^{(2)}& &\\& & \ddots &\\ & & & \tilde{r}_{12,s}^{(P)} \end{pmatrix}\hat{B}\left(h\right)\begin{pmatrix} r_{23,s}^{(1)} & & & \\ & r_{23,s}^{(2)}& &\\& & \ddots &\\ & & & r_{23,s}^{(P)} \end{pmatrix}\begin{pmatrix} \lambda_{23,s}^{(1)}\\ \lambda_{23,s}^{(2)} \\ \vdots \\ \lambda_{23,s}^{(P)}  \end{pmatrix},\\
\end{split}
\end{equation} 

Similarly, we can pick up 
\begin{eqnarray}
&\left|\bm{r}_{23}^{(k)}\right|^{2}:=\left|\bm{r}_{2}^{(k)}-\bm{r}_{3}^{(k)}\right|^{2}=l_{23}^{2} ~~\left(k=1,...,P\right),\\
&\left|\bm{r}_{31}^{(k)}\right|^{2}:=\left|\bm{r}_{3}^{(k)}-\bm{r}_{1}^{(k)}\right|^{2}=l_{31}^{2} ~~\left(k=1,...,P\right),
\end{eqnarray} 
and derive the corresponding equations.

In sum, the equations we need to solve are expressed as
\begin{equation}\label{eq.iteration}
\begin{pmatrix} \sum_{s=1}^{3}\left(\tilde{r}_{12,s}^{(1)}\right)^{2}-l_{12}^{2}\\ \vdots \\  \sum_{s=1}^{3}\left(\tilde{r}_{12,s}^{(P)}\right)^{2}-l_{12}^{2}\\
\sum_{s=1}^{3}\left(\tilde{r}_{23,s}^{(1)}\right)^{2}-l_{23}^{2} \\ \vdots \\ \sum_{s=1}^{3}\left(\tilde{r}_{23,s}^{(P)}\right)^{2}-l_{23}^{2}\\ 
\sum_{s=1}^{3}\left(\tilde{r}_{31,s}^{(1)}\right)^{2}-l_{31}^{2} \\ \vdots \\ \sum_{s=1}^{3}\left(\tilde{r}_{31,s}^{(P)}\right)^{2}-l_{31}^{2}
\end{pmatrix}=\begin{pmatrix} & J_{11} &  J_{12} &   J_{13} \\&  & & &\\& J_{21} & J_{22}& J_{23}\\& & & &\\&  J_{31} & J_{32} & J_{33}\end{pmatrix}\begin{pmatrix} \lambda_{12,s}^{(1)} \\ \vdots \\ \lambda_{12,s}^{(P)}\\ \\ \lambda_{23}^{(1)} \\ \vdots \\ \lambda_{23}^{(P)} \\ \\ \lambda_{31}^{(1)} \\ \vdots 
\\ \lambda_{31}^{(P)} 
\end{pmatrix}
\end{equation}
with
\begin{eqnarray}
&J_{i,i-1}=-\displaystyle{\sum_{s=1}^{3}\frac{2h}{m_{i}}}\begin{pmatrix} \tilde{r}_{i,i+1,s}^{(1)}  & & \\  & \ddots &\\  & & \tilde{r}_{i,i+1,s}^{(P)} \end{pmatrix}\hat{B}\left(h\right)\begin{pmatrix} r_{i-1,i,s}^{(1)} & &  \\ & \ddots &\\  & & r_{i-1,i,s}^{(P)}\end{pmatrix},\\
&J_{i,i}=\displaystyle{\sum_{s=1}^{3}\frac{2h}{m_{i,i+1}}}\begin{pmatrix} \tilde{r}_{i,i+1,s}^{(1)} & &  \\ & \ddots &\\  & & \tilde{r}_{i,i+1,s}^{(P)} \end{pmatrix}\hat{B}\left(h\right)\begin{pmatrix} r_{i,i+1,s}^{(1)} & & \\ & \ddots &\\  & & r_{i,i+1,s}^{(P)}\end{pmatrix},\\
&J_{i,i+1}=-\displaystyle{\sum_{s=1}^{3}\frac{2h}{m_{i+1}}}\begin{pmatrix} \tilde{r}_{i,i+1,s}^{(1)} & & \\  & \ddots &\\  & & \tilde{r}_{i,i+1,s}^{(P)} \end{pmatrix}\hat{B}\left(h\right)\begin{pmatrix} r_{i+1,i+2,s}^{(1)} & & \\& \ddots &\\ & & r_{i+1,i+2,s}^{(P)}\end{pmatrix},
\end{eqnarray}
and the index $i$ is subject to cyclic condition $ i= i \left(mod~3\right)$.

The next procedure is the same as in the standard SHAKE. By solving Eq.\eqref{eq.iteration}, we can get $\Lambda_{c}^{(k)}$ and update the guess $\bm{\tilde{r}}_{ij}^{(k)}$. The iteration will continue until $\left|\bm{\left|\tilde{r}}_{ij}^{(k)}\right|^{2}-l_{ij}^{2}\right|<\epsilon$ for all $k$ and a sufficiently small $\epsilon$.

The remaining part is to determine $\Lambda_{cv}^{(k)}=\left\{\sigma_{ij}^{(k)}\right\}$ and adjust the velocity to satisfy the hidden constraints \eqref{water_velocity_constraints}. Denote by 
\begin{equation}
\dot{\bm{r}}_{ij}^{(k)}=\frac{\bm{p}_{i}^{(k)}}{m_{i}}-\frac{\bm{p}_{j}^{(k)}}{m_{j}},
\end{equation}
If we pick up the constraints $\dot{\bm{r}}_{12}^{(k)} \cdot  \bm{r}_{12}^{(k)}=0$, then $\dot{\bm{r}}_{12}^{(k)}=\left(\dot{r}_{12,1}^{(k)},\dot{r}_{12,2}^{(k)},\dot{r}_{12,3}^{(k)}\right)$ satisfies
\begin{eqnarray}\label{eq.46}
\begin{split}
\begin{pmatrix} \dot{r}_{12,s}^{(1)}\left(t_{n+1}\right)\\ \dot{r}_{12,s}^{(2)}\left(t_{n+1}\right) \\ \vdots \\ \dot{r}_{12,s}^{(P)}\left(t_{n+1}\right) \end{pmatrix}= &\begin{pmatrix} \dot{\tilde{r}}_{12,s}^{(1)}\left(t_{n+1}\right)\\ \dot{\tilde{r}}_{12,s}^{(2)}\left(t_{n+1}\right) \\ \vdots \\ \dot{\tilde{r}}_{12,s}^{(P)} \left(t_{n+1}\right) \end{pmatrix}-
\frac{h}{m_{12}} \begin{pmatrix} \sigma_{12,s}^{(1)} r_{12,s}^{(1)}\left(t_{n}\right)\\\sigma_{12,s}^{(2)}  r_{12,s}^{(2)}\left(t_{n}\right) \\ \vdots \\ \sigma_{12,s}^{(P)}  r_{12,s}^{(	P)}\left(t_{n}\right)\end{pmatrix}-\frac{h}{m_{1}}\begin{pmatrix} \sigma_{31}^{(1)} r_{31,s}^{(1)}\left(t_{n}\right)\\\sigma_{31}^{(2)}  r_{31,s}^{(2)}\left(t_{n}\right) \\ \vdots \\ \sigma_{31}^{(P)}  r_{31,s}^{(P)}\left(t_{n}\right)\end{pmatrix}\\
&-\frac{h}{m_{2}} \begin{pmatrix} \sigma_{23}^{(1)} r_{23,s}^{(1)}\left(t_{n}\right)\\ \sigma_{23}^{(2)}  r_{23,s}^{(2)}\left(t_{n}\right) \\ \vdots \\ \sigma_{23}^{(P)}  r_{23,s}^{(P)}\left(t_{n}\right)\end{pmatrix},
\end{split}
\end{eqnarray} 
with $s=1,2,3$ and $\dot{\tilde{r}}_{ij,s}^{(k)}$ the initial guess of $\dot{r}_{ij,s}^{(k)}$ by putting $\sigma_{ij}^{(k)}=0$.

By multiplying Eq.\eqref{eq.46} with $\textup{diag}\left\{r_{12,s}^{(1)}\left(t_{n+1}\right), r_{12,s}^{(2)}\left(t_{n+1}\right), \cdots, r_{12,s}^{(P)}\left(t_{n+1}\right)\right\}$ and summing over $s$, we can directly solve $\sigma_{ij}^{(k)}$ without any iteration.

\begin{remark}
In the extended phase space, all the beads are connected by harmonic springs. Thus it motivates us to treat all the beads as an ensemble and use the operator $\hat{B}\left(h\right)$ to present their connection, instead of treating them independently. This provides a better way to track the highly oscillatory motions within beads.

It notes that the iterations in the trigonometric integrators are very similar to those in the MILC method\cite{BLS}, whereas the elements in the coefficient matrix $J$ are replaced by block matrices. As the Trotter number $P$ cannot be very large ($P\leq 32$) in practice, it is convenient to solve the Eq.\eqref{eq.iteration} (with a  $3P \times 3P$ coefficient matrix). Besides, the iteration convergences rapidly, like the MILC (or RATTLE) algorithm. This is because the distortion of bond length is usually very small.  

\end{remark}

\subsection{Near-conservation of the Hamiltonian}\label{sec:3.4}
It shows that the trigonometric integrator defines a flow mapping $\varphi_{h}^{\tilde{H}_{\mathcal{M}}}: \left(\bm{p}_{n},\bm{x}_{n}\right) \rightarrow \left(\bm{p}_{n+1},\bm{x}_{n+1}\right)$ associated with a perturbed Hamiltonian $\tilde{H}_{\mathcal{M}}$
\begin{equation}
\varphi^{\tilde{H}_{\mathcal{M}}}_{h}=\varphi_{h/2}^{\tilde{V}_{cv}}\circ \varphi_{h}^{\tilde{H}_{P}} \circ  \varphi_{h/2}^{\tilde{V}_{c}},
\end{equation}
subject to
\begin{align}
\begin{split}
&g \circ \varphi^{\tilde{H}_{\mathcal{M}}}_{h}=0,\\
&f \circ \varphi^{\tilde{H}_{\mathcal{M}}}_{h}=0.
\end{split}
\end{align}
$\tilde{H}_{P}$ and $\tilde{V}_{c},\tilde{V}_{cv}$ are the perturbations of the Hamiltonian functions $H_{P}$ and $V_{c}$, respectively. The numerical flow $\varphi_{h}^{\tilde{H}_{P}}$ is given by either the impulse method or the mollified impulse method. 

For simplicity, we only consider the numerical schemes with constant time step. The following theorem is based on the results of \cite{RS,CHL2}. In \cite{RS}, the author reformulated the constrained problem as an unconstrained one and thoroughly analyzed the numerical integration through the backward error analysis. Since we are more interested in the numerical stability of the trigonometric integrators with a relatively large time step $h$ (for instance, $h$ is assumed to have a lower bound $h \geq c_{0}$) ,  our proof is based on the modulated Fourier expansion\cite{CHL1,CHL2}.

To present the main theorem, we need the following assumptions\cite{HLW,CHL2}.

\begin{itemize}
\item The initial energy is bounded independent of natural frequencies of beads
\begin{equation}
H_{0}\left(\bm{p},\bm{x}\right)=\sum_{j=1}^{N}\sum_{k=1}^{P}\left[\frac{\left(p_{j}^{\left(k\right)}\right)^{2}}{2m_{j}}+\frac{m_{j}}{2\beta_{P}^{2} \hbar^{2}}\left(x_{j}^{\left(k\right)}-x_{j}^{\left(k-1\right)}\right)^{2}\right] \leq E.
\end{equation}

\item For each degree of freedom $j$, $\phi\left(h\right)\bm{x}_{j,n}$ stay in a compact subset of a domain on which the potential $V\left(\bm{x}\right)$ is smooth.

\item Let $\bm{\omega}=\left(\omega_{1},\cdots, \omega_{P-1}\right)$ and $\mathcal{K}=\left\{\bm{k} \in \mathbb{Z}^{P-1}: \bm{k} \cdot \bm{\omega}=0\right\}$, then there exists a $N$ and $c$ such that 
\begin{equation}
\left|\sin\left(h \bm{k} \cdot \bm{\omega}\right)\right|\geq c\sqrt{h}
\end{equation}
for all $\bm{k}\in \mathbb{Z}^{P-1}\setminus \mathcal{K}$ and $\left|\bm{k}\right| \leq N$. It is termed the numerical non-resonance condition.
\end{itemize}

\begin{theorem}
The trigonometric integrator $\varphi_{h}^{H_{\mathcal{\tilde{M}}}}$ is symplectic, time-reversible and constraint-preserving. Moreover, with the above assumptions and an additional condition 
\begin{equation}
\left|\phi\left(h\omega_{i}\right)\right|\leq C \left|\textup{sinc}\left(\frac{1}{2}h\omega_{i}\right)\right|~\left(i=1,\cdots,P\right)
\end{equation}
we have  
\begin{equation}
H_{P}\left(\bm{p}_{n},\bm{x}_{n}\right)=H_{P}\left(\bm{p}_{0},\bm{x}_{0}\right)+\mathcal{O}\left(h\right),
\end{equation}
otherwise
\begin{equation}
H_{P}\left(\bm{p}_{n},\bm{x}_{n}\right)=H_{P}\left(\bm{p}_{0},\bm{x}_{0}\right)+\mathcal{O}\left(h^{1/2}\right).
\end{equation}
\end{theorem}

The preservation of constraints is obvious. Since the impulse method and the mollified impulse method are time-reversible, the above integrator is also time-reversible. In addition, the symplecticity of the integrator has been proved in \cite{RS}. Therefore, we only need to prove the near-conservation of the Hamiltonian.  The detailed proof is put in the appendix.

\section{Evaluation of methods}
The numerical results are presented by making a comparisons between the trigonometric integrator and the original RATTLE algorithm. The stability of the numerical scheme with varying time step is also discussed.  It shows that the trigonometric conserves the Hamiltonian much better than the RATTLE algorithm and allows a longer time step.

\subsection{Test problem}
The simulations were performed using the SPC/E force field at 298$\textup{K}$ with a density of 0.998 $\textup{cm}^{-1}$, which had been used in simulating quantum diffusion in liquid water \cite{MMQ}. The interactions between molecular pairs are
\begin{equation}
V_{ij}=\sum_{k\in i}\sum_{k'\in j}\frac{Q_{k}Q_{k'}}{r_{kk'}} + \frac{A}{r_{\textup{OO}}^{12}}-\frac{B}{r_{\textup{OO}}^{6}},
\end{equation}
involving a Coulomb contribution and a Lennard-Jones interaction between oxygen atoms.    Parameters are listed in Table \ref{T.1}.  Since the lengths and angles of intramolecular bonds are fixed, we are not bothered by the intramolecular forces.

\begin{table}
 \centering
 \caption{\label{T.1} Parameters in SPC/E water potential.}
 \begin{tabular}{lccccl}
  \toprule
  \toprule
  Parameter & & & & & Value\\
  \midrule
  $r\left(\textup{OH}\right)\left(\textup{\AA}\right)$ &  & & & &1.0 \\
  $\angle \left(\textup{HOH}\right) \left(\textup{deg}\right)$ &  &  & & &109.47  \\
  $A\left(\textup{kJ} \cdot \textup{\AA}^{12} \cdot \textup{mol}^{-1}\right)$ &  & & & &$2.633 \times 10^{6}$\\
  $B\left(\textup{kJ} \cdot  \textup{\AA}^{6} \cdot \textup{mol}^{-1}\right)$ &  & & & &$2.617 \times 10^{3}$\\
  $Q_{\textup{O}}\left(\left|(e)\right|\right)$ & & & & & -0.8476\\
  $Q_{\textup{H}}\left(\left|(e)\right|\right)$ & & & & & 0.4238\\
  \bottomrule
  \bottomrule
  \end{tabular}
\end{table}

To truncate the intermolecular forces and split the electrostatic potential and Lennard potential into fast and slow parts smoothly, we introduce a switching function $S(r)$\cite{HFB}
\begin{equation}
S\left(r\right)= \left\{
\begin{array}{ll}
\displaystyle{1,~~\left(r<r_{c}-\Delta r\right)}\\
\displaystyle{1+R^{2}(2R-3),~~\left(r_{c}-\Delta r \leq r \leq r_{c}\right)}\\
\displaystyle{0,~~\left(r_{c}<r\right)}
\end{array}\right. 
\end{equation}
where $R=\left[r-\left(r_{c}-\Delta r\right)\right]/\Delta r$, $r$ is the interatomic distance, $r_{c}$ is the short-range cutoff and $\Delta r$ is the healing length. 

The potential energy is given by 
\begin{equation}
\begin{split}
&V^{\textup{electrostatic,fast}}= \frac{Q_{k}Q_{k^{\prime}}}{r_{k k^{\prime}}} S(r_{k k^{\prime}}),\\
&V^{\textup{electrostatic,slow}}=\frac{Q_{k}Q_{k^{\prime}}}{r_{k k^{\prime}}} \left(1-S(r_{k k^{\prime}})\right),\\
&V^{\textup{Lennard-Jones,fast}}=\left(\frac{A}{r_{\textup{OO}}^{12}}-\frac{B}{r_{\textup{OO}}^{6}}\right) S(r_{\textup{OO}}),\\
&V^{\textup{Lennard-Jones,slow}}=\left(\frac{A}{r_{\textup{OO}}^{12}}-\frac{B}{r_{\textup{OO}}^{6}}\right) \left(1-S(r_{\textup{OO}})\right).
\end{split}
\end{equation}
with $r_{ij}=\left| \bm{r}_{i}-\bm{r}_{j} \right|$. 

\subsection{Performance metrics}
Since we expected to generate trajectories with correct statistical properties, the performance metrics were based on the drift of the total Hamiltonian.

In each simulation, the history of trajectories and all the components of energy were recorded. Several quantities, devised in \cite{IRS} and \cite{HFB}, were used to measure the conservation of the ring polymer Hamiltonian using different numerical methods.  

The precent relative drift is given by $D=d/K$, where $K$ is the average kinetic energy and $d$ is the absolute energy drift (the coefficient of a linear regression model on the energy). Noise refers to the variance of linear regression model. It is reported that $D$ is a robust metric of drift in classical molecular dynamics simulations.

The absolute and relative variation in the true energy, $\Delta E$ and $\Delta E_{r}$ , are given by 
\begin{equation}
\Delta E=\frac{1}{J}\sum_{i=1}^{J}\left|E(i)-E_{initial}\right|,\quad \Delta E_{r}=\frac{1}{KJ}\sum_{i=1}^{J}\left|E(i)-E_{initial}\right|,
\end{equation}
where $E_{initial}$ is initial total energy, $E\left(i\right)$ is instantaneous total energy, and $J$ is the simulation length. $\Delta E_{r}$ can measure the distance between the true energy surface in phase space and a perturbed energy surface arising from the use of finite time step $h$.

\subsection{Numerical results}
In the subsequent simulations, the masses of oxygen and hydrogen atom were 15.999 and 1.008, respectively. Both the reduced Planck constant $\hbar$ and the Boltzmann constant $\beta$ were chosen as 1.

First, we compared the stability of the RATTLE algorithm, the trigonometric method (Impulse-R) and the trigonometric method with mollified forces (MOLLY-R).  The cell contained eight water molecules and each molecule was extended to 16 beads  (128 quasi-particles). For testing purposes, we ignored the interactions between the molecules and their periodic images in the neighboring cells. 

In order to test the stability of trigonometric methods over a long time, the time length of simulations were taken as 750, with different time stepsizes $0.02\sim0.075$. The RATTLE algorithm was also tested, with much smaller time stepsizes and the time length of simulations taken as 20. The numerical results are summarized in Table \ref{table_2}.
  \begin{table}
 \centering
 \caption{\label{table_2} Results for simulations of ring-polymer Hamiltonian dynamics with SPC/E force field at 298K. A cell contained 8 molecules and each molecule was extended to 16 beads. The interactions between molecules and their periodic images were ignored.}
 \begin{tabular}{cccccccc}
  \toprule
  \toprule
  $h$ & Method & Drift & Noise  & $\Delta E$ & $\Delta E_{r}$\\
  \midrule
  0.02 & Impulse-R  & $9.0338 \times 10^{-7}$ & $3.7599 \times 10^{-5}$ & 6.8000 & $3.6146 \times 10^{-6}$\\
  0.02 & MOLLY-R & $7.5497 \times 10^{-7}$ & $4.8081 \times 10^{-5}$ & 7.7759 & $4.1334 \times 10^{-6}$\\
  0.05 & Impulse-R  & $1.7343 \times 10^{-6}$ & $3.3363 \times 10^{-5}$ & 6.6469 & $3.5334 \times 10^{-6}$\\
  0.05 & MOLLY-R  & $1.5962 \times 10^{-6}$ & $4.0568 \times 10^{-5}$ & 7.0997 & $3.7741 \times 10^{-6}$\\
  0.075 & Impulse-R  & $7.0243 \times 10^{-7}$ & $5.2225 \times 10^{-5}$ & 7.8169 & $4.1555 \times 10^{-6}$\\
  0.075 & MOLLY-R  & $1.0121 \times 10^{-6}$ & $5.5280 \times 10^{-5}$ & 8.2969 & $4.4106 \times 10^{-6}$\\
  0.0002 & RATTLE  & $-7.7938 \times 10^{-6}$ & $2.2414 \times 10^{-5}$ & 9.3492 & $4.9692 \times 10^{-6}$\\
  0.0005 & RATTLE  & $-4.7918 \times 10^{-5}$ & $7.8758 \times 10^{-4}$ & 83.8298 & $4.4560 \times 10^{-5}$\\
    \bottomrule
  \bottomrule 
  \end{tabular}
\end{table}
  
It shows that the trigonometric methods are superior to the original RATTLE algorithm in the conservation of the Hamiltonian function.  In this simple case, the trigonometric integrators are stable even when $h\omega_{max} =2.4$ ($\omega_{max}$ is the maximal natural frequency of beads). On the contrary, a significantly small time step ($h=0.0002$) is needed in the RATTLE algorithm, and its numerical stability is contaminated when the time step grows larger ($h=0.0005$).
 
\begin{figure}[H]
    \centering
    \includegraphics[width=2.4in,height=1.8in]{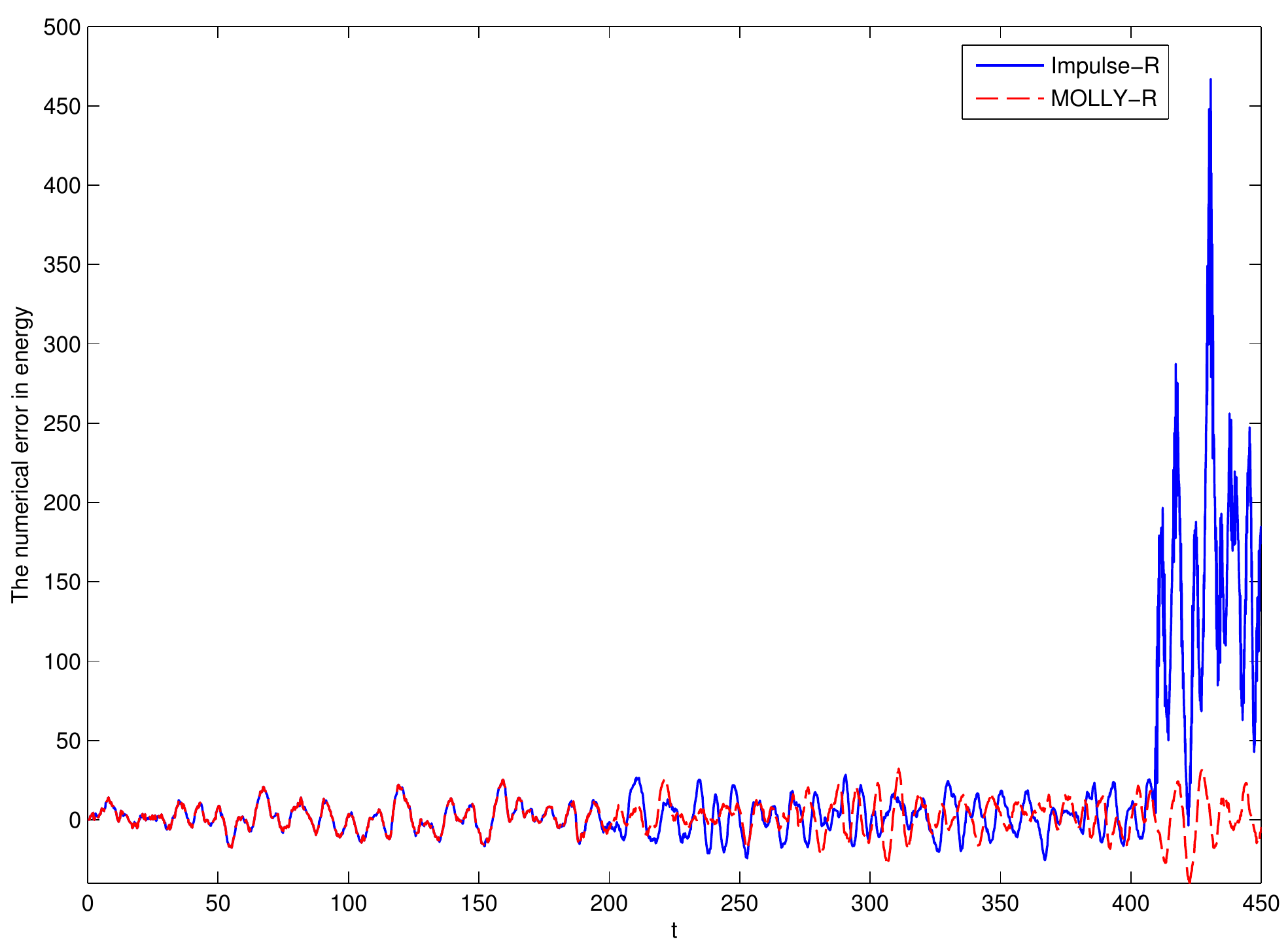}
     \caption{The nonlinear instability induced by the numerical resonance when $h=0.125$. The numerical instability is ameliorated by the mollified forces.}
     \label{fig_1}  
\end{figure}

The use of mollified forces seems to make the trigonometric integrator less stable, but it can ameliorate the numerical resonances induced by nonlinear instability\cite{MIS,HLW}. We performed the simulation with a large time step $h=0.125$ and observed that the Impulse-R was not stable and the iteration failed to convergence due to the large distortion of bond lengths, whereas the MOLLY-R didn't suffer from this problem. The numerical errors in energy ($E\left(i\right)-E_{initial}$) are plotted in Figure \ref{fig_1}.  
 
\begin{figure}[!h]
    \centering
    \subfigure[Percent relative drift $D$ vs $h$]{
    \includegraphics[width=2.4in,height=1.8in]{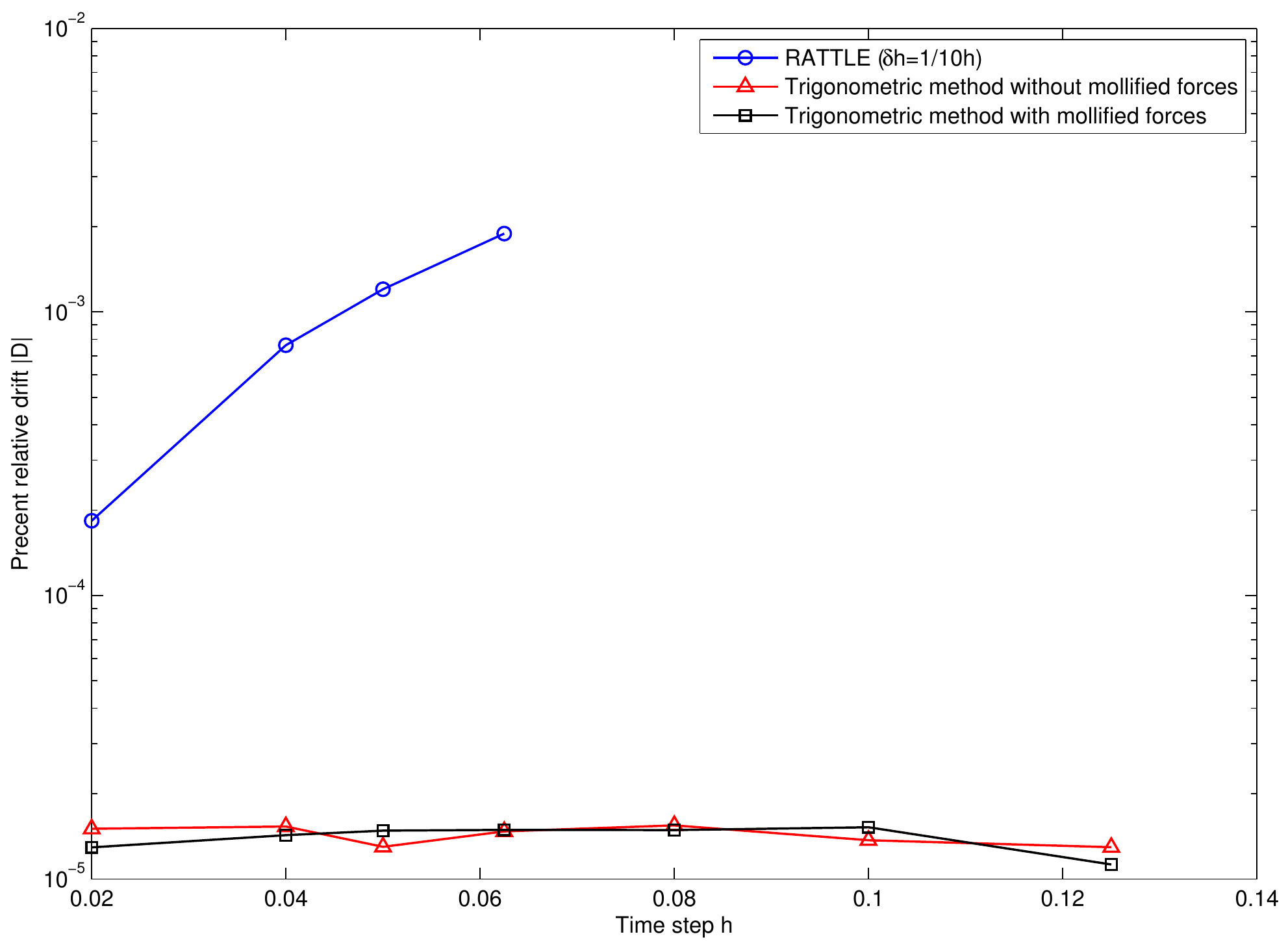}}
    \hspace{0.2in}
    \subfigure[Percent relative variation in energy $\Delta E_{r}$ ($\textup{kcal}\cdot \textup{mol}^{-1}\textup{K}^{-1}$) vs $h$]{
    \includegraphics[width=2.4in,height=1.8in]{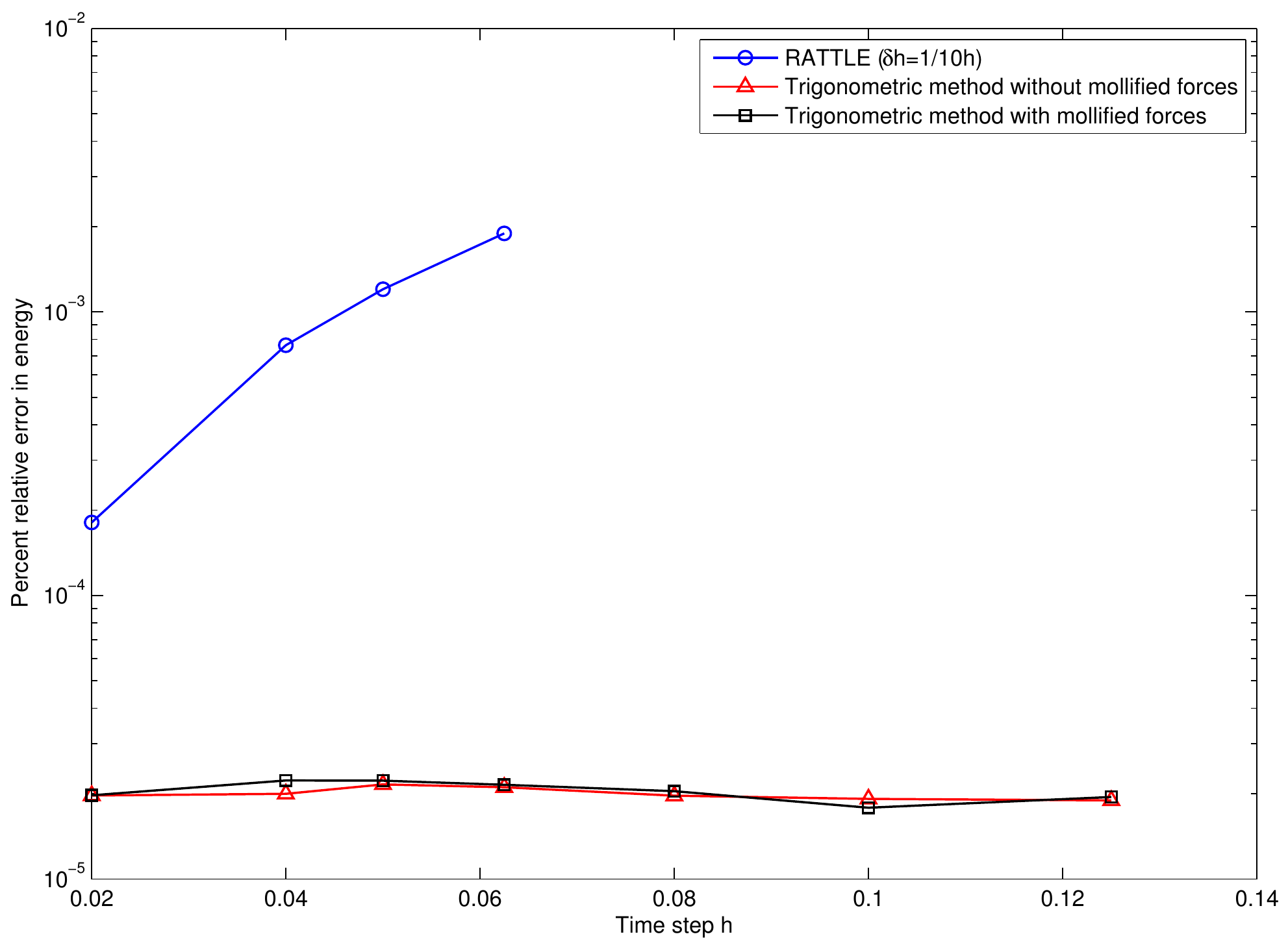}}
     \caption{The percent relative drift $D$ and variation in energy $\Delta E_{r}$ vs time step $h$.}
         \label{fig_2}  
\end{figure}

We further investigated the numerical stability of the trigonometric methods under different time steps and made a comparison between the trigonometric integrators and the velocity SHAKE-I algorithm (RATTLE using Impulse MTS as the integrator \cite{IRS}, denoted by RATTLE-I). The cell contained 8 molecules and each molecule was extended to 8 beads. The time length was 250 and the time step was chosen from $0.02$ to $0.125$. The inner time step for the velocity SHAKE-I algorithm was $\delta h=\frac{1}{10} h$. The long-range forces were truncated at the nearest neighboring cells. The numerical results are listed in Table \ref{Table_3}, and the percent relative drift $D$ and the percent relative variation in energy $\Delta E_{r}$ under different time steps are plotted in Figure \ref{fig_2} (in logarithm scale).
 \begin{table}
 \centering
 \caption{\label{Table_3} Results for simulations of ring-polymer Hamiltonian dynamics with SPC/E force field at 298K. A cell contained 8 molecules and each molecule was extended to 8 beads. Electrostatic potential and Lennard-Jones potential are truncated at the nearest neighboring cells.}
 \begin{tabular}{cccccccc}
  \toprule
  \toprule
  $h$ & Method & $\delta h$ & Drift & Noise  & $\Delta E$ & $\Delta E_{r}$\\
  \midrule
  0.02 & RATTLE-I & 0.002 &  $-1.836\times 10^{-4}$ & $5.0892\times 10^{-3}$ & 71.056 & $1.8110\times 10^{-4}$\\
  0.02 & RATTLE-I & 0.004 &  $-7.66\times 10^{-4}$ & $7.9759\times 10^{-2}$ & 300.31 & $7.66\times 10^{-4}$\\
  0.02 & Impulse-R & - &  $1.5031\times 10^{-5}$ & $1.4935\times 10^{-4}$ & 7.7355 & $1.9713\times 10^{-5}$\\
  0.02 & MOLLY-R & - &  $1.2938\times 10^{-5}$ & $1.3692\times 10^{-4}$ & 7.7442 & $1.9735\times 10^{-5}$\\
  0.04 & RATTLE-I & 0.004 &  $-7.64\times 10^{-4}$ & $7.9800\times 10^{-2}$ & 299.60 & $7.64\times 10^{-4}$\\
  0.04 & Impulse-R & - &  $1.5331\times 10^{-5}$ & $1.7126\times 10^{-4}$ & 7.8388 & $1.9976\times 10^{-5}$\\
  0.04 & MOLLY-R & - &$1.4290\times 10^{-5}$ & $1.9974\times 10^{-4}$ & 8.7430 & $2.2280\times 10^{-5}$\\
  0.05 & RATTLE-I & 0.005 &  $-1.204\times 10^{-3}$ & $1.9458\times 10^{-1}$ & 472.06 & $1.204\times 10^{-3}$\\
  0.05 & Impulse-R & - &  $1.2997\times 10^{-5}$ & $1.7242\times 10^{-4}$ & 8.4642 & $2.1570\times 10^{-5}$\\
  0.05 & MOLLY-R & - &$1.4820\times 10^{-5}$ & $1.9770\times 10^{-4}$ & 8.7246 & $2.2233\times 10^{-5}$\\
  0.0625 & RATTLE-I & 0.00625 &  $-1.891\times 10^{-3}$ & $4.7540\times 10^{-1}$ & 741.39 & $1.893\times 10^{-3}$\\
  0.0625 & Impulse-R & - &  $1.4716\times 10^{-5}$ & $1.7006\times 10^{-4}$ & 8.2727 & $2.1082\times 10^{-5}$\\
  0.0625 & MOLLY-R & - &$1.4916\times 10^{-5}$ & $1.9367\times 10^{-4}$ & 8.4417 & $2.1512\times 10^{-5}$\\
  0.08 & Impulse-R & - &  $1.5470\times 10^{-5}$ & $1.6051\times 10^{-4}$ & 7.7520 & $1.9686\times 10^{-5}$\\
  0.08 & MOLLY-R & - &$1.4877\times 10^{-5}$ & $1.6342\times 10^{-4}$ & 8.0114 & $2.0416\times 10^{-5}$\\
  0.1 & Impulse-R & - &  $1.3702\times 10^{-5}$ & $1.2784\times 10^{-4}$ & 7.5216 & $1.9168\times 10^{-5}$\\
  0.1 & MOLLY-R & - &$1.5237\times 10^{-5}$ & $1.1066\times 10^{-4}$ & 7.0072 & $1.7857\times 10^{-5}$\\
  0.125 & Impulse-R & - &  $1.2950\times 10^{-5}$ & $1.2542\times 10^{-4}$ & 7.4286 & $1.8930\times 10^{-5}$\\
  0.125 & MOLLY-R & - &$1.1248\times 10^{-5}$ & $1.6532\times 10^{-4}$ & 7.6385 & $1.9465\times 10^{-5}$\\
  \bottomrule
  \bottomrule
  \end{tabular}
\end{table}

For the trigonometric integrators, both $D$ and $\Delta E_{r}$ stay a relatively small level with some fluctuations. It is observed that both methods preserve the Hamiltonian accurately even under a very large time step, since they solve the harmonic oscillations within beads exactly. Besides, the Impulse-R method (without mollified forces) is more stable than the MOLLY-R method in general. On the contrary, the RATTLE-I method gives far less accurate numerical results, and both $D$ and $\Delta E_{r}$ nearly grow exponentially when $h$ increases.  

Finally, we examined the numerical stability of the trigonometric integrators with varying time steps. We considered a cell containing 27 molecules and extended each one into 4 beads. The time length of the simulations was 250, with time step $h$ from $0.025$ to $0.1$. The long-range forces were truncated at  8\AA, and the healing distance was $\Delta r=4.5$\AA. 

The numerical results are presented in Table \ref{Table_4}. It shows that the trigonometric integrator with varying time step is also stable to some extent. Thus it may facilitate the inclusion of full electrostatic forces and Lennard-Jones interactions using the Ewald summation\cite{EPB}.  

\begin{table}
 \centering
 \caption{\label{Table_4} Results for simulations of ring-polymer Hamiltonian dynamics with SPC/E potential at 298K. A cell contained 27 molecules and each molecule is extended to 4 beads. Electrostatic potential and Lennard-Jones potential are truncated at 8\AA, with the healing distance $\Delta r=4.5$\AA.}
 \begin{tabular}{cccccccc}
  \toprule
  \toprule
  $h$ & Method & $\delta h$ & $\Delta r$ & Drift & Noise  & $\Delta E$ & $\Delta E_{r}$\\
  \midrule
  0.025 & Impulse-R & - & - & $-4.55\times 10^{-5}$ & $5.71 \times 10^{-3}$ & 6.7813 & $5.14\times 10^{-4}$\\
  0.05 & Impulse-R & - & -  & $-7.54\times 10^{-5}$ & $4.31 \times 10^{-3}$ & 5.9423 & $4.51\times 10^{-4}$\\
  0.1 & Impulse-R & - & - & $-2.46\times 10^{-5}$ & $4.73 \times 10^{-3}$ & 6.2156 & $4.73\times 10^{-4}$\\
  0.1 & MOLLY-R & 0.05 & 4.5 & $-6.26\times 10^{-5}$ & $3.73 \times 10^{-3}$ & 5.7808 & $4.39\times 10^{-4}$\\
  0.1 & Impulse-R & 0.05 & 4.5  & $-2.33\times 10^{-5}$ & $5.21 \times 10^{-3}$ & 6.3992 & $4.86\times 10^{-4}$\\
  0.1 & MOLLY-R & 0.025 & 4.5 & $-1.55\times 10^{-5}$ & $5.71 \times 10^{-3}$ & 7.4740 & $5.67\times 10^{-4}$\\
  0.1 & Impulse-R & 0.025 & 4.5 & $ -3.49\times 10^{-5}$ & $5.57 \times 10^{-3}$ & 6.6911 & $5.07\times 10^{-4}$\\
  \bottomrule
  \bottomrule
  \end{tabular}
\end{table}

\begin{figure}[H]
    \centering
    \includegraphics[width=2.4in,height=1.8in]{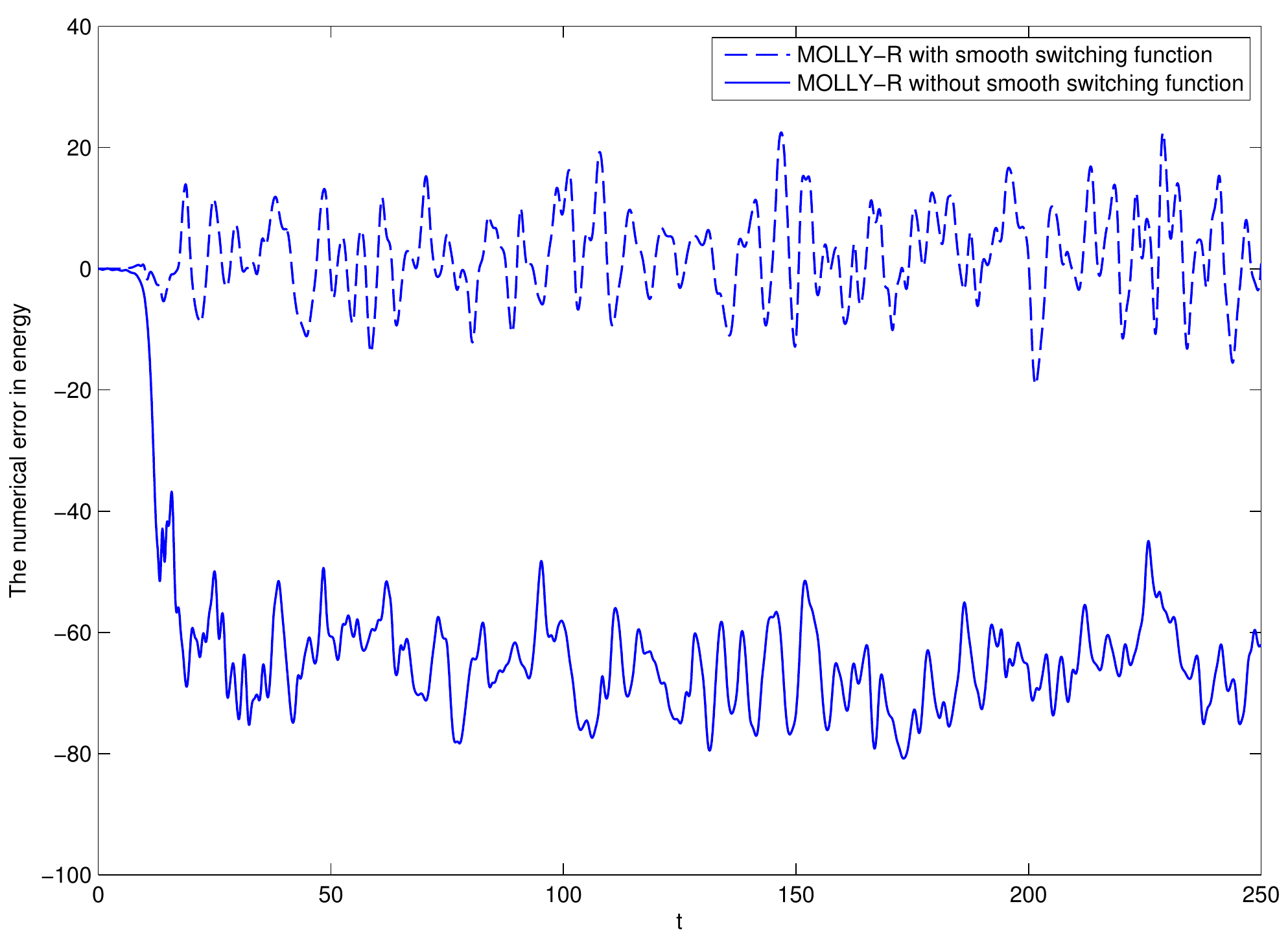}
     \caption{The numerical instability created by non-smooth switch function.}
     \label{fig_3}  
\end{figure}

The instability of MOLLY-R was observed when a non-smooth truncation function was used. For instance, we split $V^{\textup{nonbond}}$ into two parts,
\begin{equation}
V^{\textup{nonbond}}=V^{\textup{nonbond}}\bm{1}_{\left[r>r_{h} \right]}+V^{\textup{nonbond}}\bm{1}_{\left[r\leq r_{h} \right]},
\end{equation} 
The indicator function $\bm{1}_{\left[r>r_{h} \right]}$ is clearly not a smooth function.  It shows in Figure \ref{fig_3} that the energy cannot be conserved when a direct truncation is used. Actually, the trajectories become incorrect after a short time. 

\section{Conclusion}
In this paper we discuss a class of symplectic and time-reversible numerical integrators for the constrained ring polymer Hamiltonian system. The integrators are formulated via the composition of subflows and make full use of normal mode representation, thereby achieving a better conservation of the Hamiltonian in the extended phase space. We analyze the near-conservation of the Hamiltonian in the framework of modulated Fourier expansion, and present the numerical accuracy by simulating a water model with extended SPC/E force field.

Although we mainly focus on the numerical integration of constrained Hamiltonian dynamics, our method is expected to be applicable for the generalized Langevin dynamics with constraints. For instance, we can formulate the integrator by compositing the stochastic force 
\begin{equation}
\begin{split}
&e^{-h L}\simeq e^{-(h/2) L_{\gamma}} e^{-h L_{H}} e^{-(h/2) L_{\gamma}},\\
&e^{-h L}\simeq e^{-(h/2) L_{NHC}} e^{-h L_{H}} e^{-(h/2) L_{NHC}},\\
\end{split}
\end{equation}
where $L_{\gamma}$, $L_{NHC}$ and $L_{H}$ are the Liouvillian of Fokker-Planck equation, Nos\'{e}-Hoover chain and ring polymer Hamiltonian system, respectively \cite{CPMM,BPA}. Thus, the Langevin thermostatting and Nos\'{e}-Hoover-like thermostatting can also be performed using the trigonometric methods. In the future work, we would like to examine and analyze the stability of the trigonometric integrator in the generalized Langevin dynamics.\\

\hspace{-0.5cm}\textbf{Acknowledgements}\quad The author would like to thank Professor Tao Wu in the Department of Chemistry, Zhejiang University, for discussions on the quantum chemistry and the theory of ring-polymer molecular dynamics. 

\section*{Appendix}
In this section, we discuss the proof of the main theorem in the Section \ref{sec:3.4}.  It begins by taking canonical transform
\begin{equation}
\tilde{\bm{p}}_{j}=\frac{1}{\sqrt{m_{j}}}U\bm{p}_{j},\quad \tilde{\bm{x}}_{j}=\sqrt{m_{j}}U\bm{x}_{j},
\end{equation}
and denote by
\begin{align}
&\tilde{V}\left(\tilde{\bm{x}}\right)=\sum_{k=1}^{P}V\left(x_{1}^{(k)},\cdots, x_{N}^{(k)}\right),\\
&\frac{1}{2\epsilon}\tilde{g}\left(\tilde{\bm{x}}\right)^{\tau}\tilde{g}\left(\tilde{\bm{x}}\right)=\sum_{k=1}^{P}\frac{1}{2\epsilon} g\left(\bm{x}^{(k)}\right)^{\tau}g\left(\bm{x}^{(k)}\right).
\end{align}
then it yields a Hamiltonian with the form (we drop the tilde for brevity)
\begin{equation}\label{Eq.61}
H_{\mathcal{M}}\left(\dot{\bm{x}},\bm{x}\right)=\frac{1}{2}\sum_{j=1}^{N}\dot{\bm{x}}_{j}^{\tau}\dot{\bm{x}}_{j}+\frac{1}{2}\sum_{j=1}^{N}\bm{x}_{j}^{\tau} \Omega^{2}\bm{x}_{j}+V\left(\bm{x}\right)+g\left(\bm{x}\right)^{\tau} \Lambda.
\end{equation}

Recall that $\bm{x}_{j,n}=\left(x_{j}^{(1)}\left(t_{n}\right),\cdots, x_{j}^{(P)}\left(t_{n}\right)\right)^{\tau}$, then the trigonometric integrator $\varphi_{h}^{\tilde{H}_\mathcal{M}}$ can be written in two-step form
\begin{equation}
\bm{x}_{j,n+1}-2\cos \left(h\Omega\right) \bm{x}_{j,n}+\bm{x}_{j,n-1}=h^{2}\textup{sinc}\left(h\Omega\right)\left(\phi\left(h\right)\bm{u}_{j,n}+\frac{1}{2}\bm{v}_{j,n}^{\tau}\left(\bm{\Lambda}_{j,n}^{c}+\bm{\Lambda}_{j,n}^{cv}\right)\right).
\end{equation}
Furthermore, since $\bm{x}_{j,n+1}=e^{hD}\bm{x}_{j,n}$, the left side can be abbreviated as
\begin{equation}
L\left(hD\right)\bm{x}_{j,n}:=\left(e^{hD}-2\cos \left(h\Omega\right)+e^{-hD}\right)\bm{x}_{j,n}.
\end{equation}

With the assumptions in Section \ref{sec:3.4}, the numerical solution of the Hamiltonian system \eqref{Eq.61} can formally admit an expansion
\begin{equation}\label{modulated_Fourier_expansion}
\bm{x}_{j,n}=\bm{y}_{j}\left(t\right)+\sum_{\bm{k} \in \mathcal{N}}e^{i \bm{k}\cdot \bm{\omega} t} \bm{z}_{j}^{k}\left(t\right)+\textup{sinc}\left(h\Omega\right) \cdot \mathcal{O}\left(t^{2}h^{N}\right),
\end{equation}
where $t=nh$, $\mathcal{N}=\left\{\bm{k} \in \mathcal{K}; \left|\bm{k}\right|<N, \bm{k} \neq \bm{0}\right\}$.

The next procedure is to construct the perturbed Hamiltonian functions $\tilde{H}_{\mathcal{M}}$ from the modulated functions $\left(\bm{z}_{j}^{-N+1}, \cdots, \bm{z}_{j}^{-1}, \bm{y}_{j}, \bm{z}_{j}^{1}, \cdots, \bm{z}_{j}^{N-1}\right)$. Substitute Eq.\eqref{modulated_Fourier_expansion} into Eq.\eqref{Eq.61} and expand $\bm{u}_{j,n}$ and $\bm{v}_{j,n}$ into Taylor series, then we compare the coefficients and obtain
\begin{align}\label{eq.65}
\begin{split}
L\left(hD\right)\bm{y}_{j} = &\textup{sinc}\left(h\Omega\right)\phi\left(h\right)\left[\phi\left(h\right)\bm{u}_{j}\left(\phi\left(h\right)\bm{z}\right)+\sum_{s\left(\alpha\right)\sim 0} \frac{1}{m!}\bm{u}^{(m)}_{j}\left(\phi\bm{z}^{(0)}\right)\left(\phi\bm{z}\right)^{\alpha}\right]\\
&+\frac{1}{2}h^{2}\textup{sinc}\left(h\Omega\right)\bm{v}_{j}\left(\bm{z}^{(0)}\right)^{\tau}\left(\bm{\Lambda}_{j,n}^{c}+\bm{\Lambda}_{j,n}^{cv}\right),
\end{split}
\end{align}
and 
\begin{align}\label{eq.66}
L\left(hD+ih \bm{k}\cdot \bm{\omega}\right)\bm{z}_{j}^{k}=h^{2}\textup{sinc}\left(h\Omega\right)\left[\sum_{s\left(\alpha\right)\sim k} \frac{1}{m!}\bm{u}^{(m)}_{j}\left(\phi\bm{y}\right)\left(\phi\bm{z}\right)^{\alpha}+\frac{1}{2}\left(\bm{z}^{k}_{j}\right)^{\tau}\left(\bm{\Lambda}_{j,n}^{c}+\bm{\Lambda}_{j,n}^{cv}\right)\right],
\end{align}
with
\begin{align}
&\bm{u}_{j}^{(m)}\left(\phi \bm{y}\right)=\bm{u}_{j}^{(m)}\left(\phi\left(h\right)\bm{y}_{1}, \cdots, \phi\left(h\right)\bm{y}_{N}\right),
&\phi\bm{z}^{\alpha}=\left(\phi\left(h\right)\bm{z}_{j}^{\alpha_{1}}, \cdots, \phi\left(h\right)\bm{z}_{j}^{\alpha_{P}}\right)
\end{align}
and $s\left(\alpha\right)=\sum_{i=1}^{P}\alpha_{P}$ satisfies the relation $s\left(\alpha\right)\sim k$, that is, $s\left(\alpha\right)-k \in \mathcal{M}$. The Taylor expansion of $\bm{v}_{j,n}$ is simplified when $g\left(\bm{x}\right)=0$ are quadratic constraints, namely, $\bm{v}_{j,n}$ are linear functions.

We multiply Eq.\eqref{eq.65} and Eq. \eqref{eq.66} by $\dot{\bm{y}}_{j}$ and $\dot{\bm{z}}_{j}^{-k}-i \bm{k}\cdot \bm{\omega} \bm{z}_{j}^{-k}$, respectively, then sum over all $\bm{k} \in \mathcal{N}$ and $j$ to obtain
\begin{align}\label{eq.68}
\begin{split}
\mathcal{O}\left(h^{N}\right)=&\sum_{j=1}^{N} \left(\dot{\bm{y}}_{j}\right)h^{-2}\textup{sinc}\left(h\Omega\right)^{-1}L\left(hD\right)\bm{y}_{j}\\
&+\sum_{j=1}^{N} \sum_{k \in \mathcal{N}}\left(\dot{\bm{z}}_{j}^{-k}-i \bm{k}\cdot \bm{\omega} \bm{z}_{j}^{-k}\right)h^{-2}\textup{sinc}\left(h\Omega\right)^{-1}L\left(hD+ih \bm{k}\cdot \bm{\omega}\right)\bm{z}_{j}^{k}\\
&+\frac{d}{dt}V\left(\bm{z}\right)+\frac{1}{2}\frac{d}{dt} g\left(\bm{z}\right)^{\tau} \left( \Lambda_{n}^{c}+\Lambda_{n}^{cv}\right),
\end{split}
\end{align}
where the right hand side can be written as a total derivate of a function $\mathcal{H}\left[\bm{z}\right]\left(t\right)$\cite{HLW}, namely,
\begin{equation}
\frac{d}{dt}\mathcal{H}\left[\bm{z}\right]\left(t\right)=\mathcal{O}\left(h^{N}\right)
\end{equation}

The following lemma presents the relation between $H_{P}\left(\dot{\bm{x}}\left(t\right), \bm{x}\left(t\right)\right)$ and $\mathcal{H}\left[\bm{z}\right]\left(t\right)$.

\begin{lemma}Under the assumptions in Section \ref{sec:3.4}, we have 
\begin{equation}
\mathcal{H}\left[\bm{z}\right]\left(t\right)=\mathcal{H}\left[\bm{z}\right]\left(0\right)+\mathcal{O}\left(th^{N}\right).
\end{equation} 
Moreover, at $t=nh$, we have
\begin{equation}
\mathcal{H}\left[z\right]\left(t\right)=H_{P}\left(\dot{\bm{x}}, \bm{x}\right)+\frac{1}{2}g\left(\bm{x}\right)^{\tau}\left(\Lambda_{n}^{c}+\Lambda_{n}^{cv}\right)+\mathcal{O}\left(h^{\nu}\right).
\end{equation}
$\nu=1$ if $\left|\phi\left(h\omega_{i}\right)\right|\leq C \left|\textup{sinc}\left(\frac{1}{2}h\omega_{i}\right)\right|$, otherwise $\nu=\frac{1}{2}$.
\end{lemma}
 
The proof of the above lemma is found in \cite{CHL2}. Actually, it gives the perturbed Hamiltonian $\tilde{H}_{\mathcal{M}}$ associated with the trigonometric integrator $\varphi_{h}^{\tilde{H}_\mathcal{M}}$ 
\begin{equation}\label{eq.71}
\tilde{H}_{\mathcal{M}}\left(\dot{\bm{x}}, \bm{x}\right)=H_{P}\left(\dot{\bm{x}}, \bm{x}\right)+\frac{1}{2}g\left(\bm{x}\right)^{\tau}\left(\Lambda_{n}^{c}+\Lambda_{n}^{cv}\right)+\mathcal{O}\left(h^{\nu}\right)\\
\end{equation}

Now we define a Poisson bracket for sufficiently smooth functions $F: \mathbb{R}^{2NP} \to \mathbb{R}^{l}$ and $G: \mathbb{R}^{2NP} \to \mathbb{R}^{k}$,
\begin{equation}
\left\{F,G\right\}=\frac{dF}{d\bm{x}} \left(\frac{dG}{d\bm{p}}\right)^{\tau}-\frac{dF}{d\bm{p}} \left(\frac{dG}{d\bm{x}}\right)^{\tau}
\end{equation}
 
 Since $f \circ \varphi^{\tilde{H}_{\mathcal{M}}}_{h}=0$ is equivalent to $\left\{f, \tilde{H}_{\mathcal{M}}\right\}=0$\cite{RS},  it yields
 \begin{equation}
 \begin{split}
 0=\left\{f, \tilde{H}_{\mathcal{M}}\right\}&=\left\{f, H_{\mathcal{M}}+g\left(\bm{x}\right)^{\tau}\left[\frac{1}{2}\left(\Lambda_{n}^{c}+\Lambda_{n}^{cv}\right)- \Lambda\right]+\mathcal{O}\left(h^{\nu}\right)\right\}\\
 &=\left\{f, H_{\mathcal{M}}\right\}+\left\{f,g\right\}\left[\frac{1}{2}\left(\Lambda_{n}^{c}+\Lambda_{n}^{cv}\right)- \Lambda\right]+ \mathcal{O}\left(h^{\nu}\right),
 \end{split}
 \end{equation}
which implies
\begin{equation}\label{eq.74}
\frac{1}{2}\left(\Lambda_{n}^{c}+\Lambda_{n}^{cv}\right)= \Lambda+\mathcal{O}\left(h^{\nu}\right).
\end{equation}

Substitute Eq.\eqref{eq.74} into Eq.\eqref{eq.71}, we have
\begin{equation}
\tilde{H}_{\mathcal{M}}=H_{\mathcal{M}}+\mathcal{O}\left(h^{\nu}\right)
\end{equation} 

Finally, since $g\left(\bm{x}\right)=\bm{0}$ at every discrete time $t=nh$, it yields
\begin{equation}
H_{P}\left(\dot{\bm{x}}_{n}, \bm{x}_{n}\right)=H_{P}\left(\dot{\bm{x}}_{0}, \bm{x}_{0}\right)+\mathcal{O}\left(h^{\nu}\right)
\end{equation}
for $t=nh\leq T$ and a bounded $T$.


\end{document}